\documentclass[journal]{IEEEtran}
\usepackage{amsmath}
\usepackage{amsthm}

\usepackage{cancel}

\usepackage{amssymb}
\usepackage{amsfonts}
\usepackage{utfsym} 
\usepackage{hyperref} 
\usepackage{cleveref}
\crefname{section}{Sec.}{Secs.}
\Crefname{section}{Section}{Sections}
\Crefname{table}{Table}{Tables}
\crefname{table}{Tab.}{Tabs.}
\Crefname{figure}{Figure}{Figures}
\crefname{figure}{Fig.}{Figs.}
\Crefname{equation}{Equation}{Equations}
\crefname{equation}{Eq.}{Eqs.}
\usepackage{url} 
\usepackage{cite} 
\usepackage{graphicx}
\usepackage{multirow}
\usepackage{makecell}
\usepackage[flushleft]{threeparttable}
\usepackage{colortbl}
\usepackage{booktabs}
\usepackage{caption}
\usepackage{subcaption}
\usepackage{xspace}
\usepackage[dvipsnames]{xcolor}
\usepackage{enumitem}
\usepackage{algorithm}
\usepackage{algpseudocode} 

\def\eg{\emph{e.g.,}\xspace}

\def\ie{\emph{i.e.,}\xspace}
\def\etal{\emph{et al.}\xspace}
\def\vs{\emph{vs.}\xspace}

\def\aka{\emph{a.k.a.}\xspace}

\newcommand{\one}{({\em i})\xspace}
\newcommand{\two}{({\em ii})\xspace}
\newcommand{\three}{({\em iii})\xspace}

\newcommand{\labelicon}{
    \tikz[baseline=-1ex]\draw[thick, fill=white] (0,0) circle (4pt)
        node {\scriptsize L};
}
\newcommand{\embeddingicon}{
    \tikz[baseline=-1ex]\draw[thick, fill=gray!30] (0,0) circle (4pt)
        node {\scriptsize E};
}
\newcommand{\imageicon}{
    \tikz[baseline=0.2ex]{
        \draw[thick, fill=blue!20] (0,0) rectangle (0.3,0.25);
        \draw (0.1,0) -- (0.1,0.25);
        \draw (0.2,0) -- (0.2,0.25);
        \draw (0,0.125) -- (0.3,0.125);
    }
}

\newcommand{\accsize}{7pt}
\newcommand{\wbicon}{
  \tikz[baseline=0.2ex]{\draw[thick] (0,0) rectangle (\accsize,\accsize);}%
}
\newcommand{\bbicon}{
  \tikz[baseline=0.2ex]{\fill[black] (0,0) rectangle (\accsize,\accsize);
                         \draw[thick] (0,0) rectangle (\accsize,\accsize);}%
}
\newcommand{\bothicon}{
    \wbicon\ \&\ \bbicon
}

\newcommand{\grayscaleicon}{
  \tikz[baseline=0.2ex]{%
    \fill[gray!50] (0,0) rectangle (0.25,0.25);
    \draw[thick] (0,0) rectangle (0.25,0.25);
  }%
}
\newcommand{\rgbicon}{
  \tikz[baseline=0.2ex]{%
    \fill[red]   (0,0) rectangle (0.0833,0.25);
    \fill[green] (0.0833,0) rectangle (0.1666,0.25);
    \fill[blue]  (0.1666,0) rectangle (0.25,0.25);
    \draw[thick] (0,0) rectangle (0.25,0.25);
  }%
}

\ifCLASSINFOpdf
\else
\fi
\begin{document}
%
\title{DiffMI: Breaking Face Recognition Privacy via Diffusion-Driven Training-Free Model Inversion}
%
%
%

\author{Hanrui Wang,~\IEEEmembership{Member,~IEEE,}
        Shuo Wang,~\IEEEmembership{Senior~Member,~IEEE,}\\
        Chun-Shien Lu,~\IEEEmembership{Member,~IEEE,}
        and Isao Echizen,~\IEEEmembership{Senior~Member,~IEEE}
\thanks{Hanrui Wang (corresponding author) and Isao Echizen are with Echizen Lab, National Institute of Informatics (NII), Tokyo, Japan, e-mail: \{hanrui\_wang, iechizen\}@nii.ac.jp. Shuo Wang is with Shanghai Jiao Tong University, Shanghai, China, e-mail: wangshuosj@sjtu.edu.cn. Chun-Shien Lu is with Institute of Information Science, Academia Sinica, Taipei, Taiwan. e-mail: lcs@iis.sinica.edu.tw.}
\thanks{This work was partially supported by JSPS KAKENHI Grants JP21H04907 and JP24H00732, by JST CREST Grant JPMJCR20D3 and JPMJCR2562 including AIP challenge program, by JST AIP Acceleration Grant JPMJCR24U3, and by JST K Program Grant JPMJKP24C2 Japan.}
}

%
%

\markboth{IEEE Transactions on Information Forensics and Security,~Vol.~21, Month~2026}%
{Wang \MakeLowercase{\textit{et al.}}: DiffMI}
%



\maketitle

\begin{abstract}
Face recognition poses serious privacy risks due to its reliance on sensitive and immutable biometric data. While modern systems mitigate privacy risks by mapping facial images to embeddings (commonly regarded as privacy-preserving), model inversion attacks reveal that identity information can still be recovered, exposing critical vulnerabilities. However, existing attacks are often computationally expensive and lack generalization, especially those requiring target-specific training. Even training-free approaches suffer from limited identity controllability, hindering faithful reconstruction of nuanced or unseen identities. In this work, we propose DiffMI, the first diffusion-driven, training-free model inversion attack. DiffMI introduces a novel pipeline combining robust latent code initialization, a ranked adversarial refinement strategy, and a statistically grounded, confidence-aware optimization objective. DiffMI applies directly to unseen target identities and face recognition models, offering greater adaptability than training-dependent approaches while significantly reducing computational overhead. Our method achieves 84.42\%--92.87\% attack success rates against inversion-resilient systems and outperforms the best prior training-free GAN-based approach by 4.01\%--9.82\%. The implementation is available at \url{https://github.com/azrealwang/DiffMI}.
\end{abstract}


%
\IEEEpeerreviewmaketitle

\section{Introduction}
\label{intro}
\begin{figure}[!t]
    \centering
    \includegraphics[width=3in]{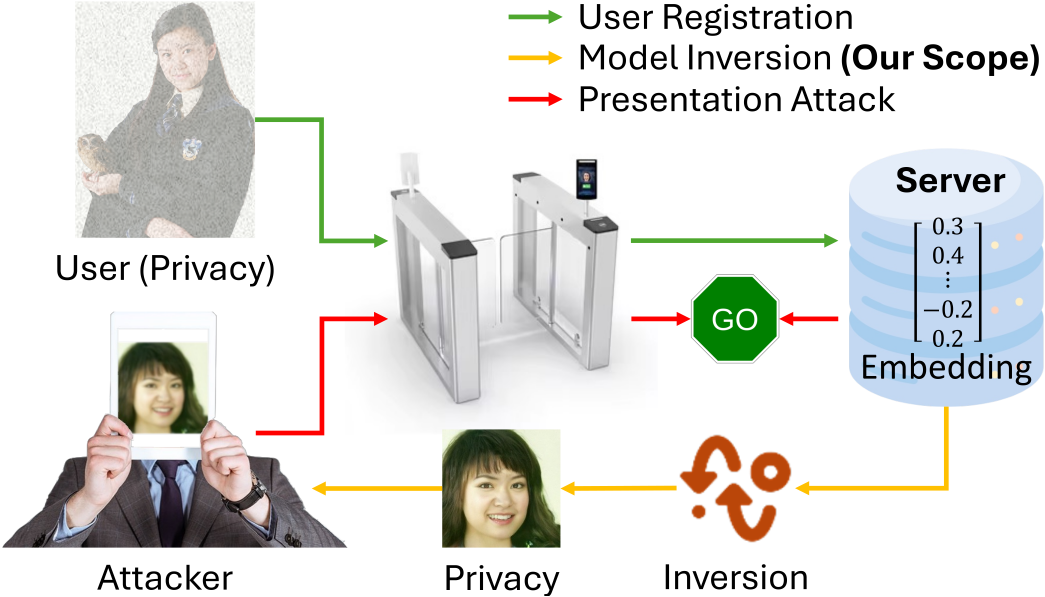}
    \caption{The threat of model inversion against embedding-based face recognition systems. Although such systems store embeddings instead of raw images for privacy protection, model inversion attacks can reconstruct facial images directly from embeddings, enabling downstream threats such as presentation attacks that can bypass authentication.}
    \label{fig_mia}
\end{figure}

Face recognition systems pose significant privacy risks due to their reliance on immutable biometric data. Large-scale leaks have occurred repeatedly, \eg U.S. CBP Cybersecurity Incident~\cite{DHSOIG_2020_CBP}, U.K. Biostar 2 breach~\cite{Taylor_2019_Guardian_BioStar2_Breach}, Shanghai Police database leak (1B records)~\cite{Fortune_2022_Shanghai_Leak}, Australia Outabox~\cite{Burt_2024_BiometricUpdate_Outabox}, and India police/military recruitment contractor leak (500GB)~\cite{Thomas_2024_CSOOnline_IndiaLeak}. \textbf{As biometric data cannot be reissued like passwords, such breaches are irreversible.} To mitigate these risks, modern systems adopt system-specific feature embeddings, improving unlinkability, scalability, generalization to unseen identities, and retrieval efficiency~\cite{MicrosoftAzureFaceAPI,AmazonRekognition,GoogleCloudVisionAPI,FacePlusPlusAPI}. As illustrated in \cref{fig_mia}, facial images are encoded into system-specific embeddings and compared using metrics such as cosine similarity or Euclidean distance. This approach was once considered privacy-preserving, as it obscures raw biometric information and prevents embeddings from being used across systems~\cite{schroff2015facenet,deng2019arcface,ji2022privacy,mi2023privacy}. A widely adopted evaluation protocol for privacy risk measures whether leaked info reveals transferable identity information, which would constitute a privacy breach~\cite{zhang2024validating,shahreza2024vulnerability,otroshi2023face,shahreza2024template,dong2023Reconstruct}.

\begin{table*}[!t]
    \footnotesize
    \centering
    \begin{threeparttable}
        \caption{Representative model inversion attacks, highlighting both influential early methods and recent advances.}
        \label{tab_relatedworks}
        \setlength{\tabcolsep}{2.8mm}{\begin{tabular}{l|cccc|c|lc|c}
            \toprule
            \multirow{2}{*}{Method}&\multicolumn{4}{c|}{Attack Cost}&Task&\multicolumn{2}{c|}{Visual Fidelity}&\multirow{2}{*}{Code}\\
            &Generator&Training-Free&Reference$^1$&Threat Model$^2$&Open-Set&Image Size$^3$&Selfies&\\
            \midrule
            MIA~\cite{fredrikson2015model} [2015]&None&$\usym{2717}$&\labelicon&\bothicon&$\usym{2717}$&\grayscaleicon&$\usym{2717}$&$\usym{2713}$\\
            NbNet~\cite{mai2018reconstruction} [2018]&DeconvNet&$\usym{2717}$&\embeddingicon&\bbicon&$\usym{2713}$&\rgbicon\ 160&$\usym{2717}$&$\usym{2713}$\\
            Amplified-MIA~\cite{zhang2023analysis} [2023]&DeconvNet&$\usym{2717}$&\labelicon&\bbicon&$\usym{2717}$&\grayscaleicon\ 64&$\usym{2717}$&$\usym{2713}$\\
            DSCasConv~\cite{shahreza2024vulnerability} [2024]\textcolor{red}{$^*$}&DeconvNet&$\usym{2717}$&\embeddingicon&\wbicon&$\usym{2713}$&\rgbicon\ 112&$\usym{2717}$&$\usym{2713}$\\
            DiBiGAN~\cite{duong2020vec2face} [2020]&C-GAN&$\usym{2717}$&\embeddingicon&\bothicon&$\usym{2713}$&\rgbicon&$\usym{2717}$&$\usym{2717}$\\
            GMI~\cite{zhang2020secret} [2020]&C-GAN&$\usym{2717}$&\labelicon&\wbicon&$\usym{2717}$&\rgbicon\ 64&$\usym{2717}$&$\usym{2713}$\\
            $\alpha$-GAN~\cite{khosravy2022model} [2022]&C-GAN&$\usym{2717}$&\labelicon&\wbicon&$\usym{2717}$&\grayscaleicon&$\usym{2717}$&$\usym{2717}$\\
            PLG-MI~\cite{yuan2023pseudo} [2023]&C-GAN&$\usym{2717}$&\labelicon&\wbicon&$\usym{2717}$&\rgbicon\ 64&$\usym{2717}$&$\usym{2713}$\\
            LOKT~\cite{nguyen2023label} [2023]&C-GAN&$\usym{2717}$&\labelicon&\bbicon&$\usym{2717}$&\rgbicon\ 128&$\usym{2717}$&$\usym{2713}$\\
            ABE-MI~\cite{wu2025label} [2025]&C-GAN&$\usym{2717}$&\labelicon&\bbicon&$\usym{2717}$&\rgbicon\ 128&$\usym{2717}$&$\usym{2717}$\\
            ID3PM~\cite{kansy2023controllable} [2023]&C-Diffusion&$\usym{2717}$&\embeddingicon&\bbicon&$\usym{2713}$&\rgbicon\ 64&$\usym{2713}$&$\usym{2717}$\\
            CDM~\cite{liu2024unstoppable} [2024]&C-Diffusion&$\usym{2717}$&\labelicon&\bbicon&$\usym{2717}$&\rgbicon\ 64&$\usym{2717}$&$\usym{2717}$\\
            C-Diff-MI~\cite{li2024model} [2024]&C-Diffusion&$\usym{2717}$&\labelicon&\wbicon&$\usym{2717}$&\rgbicon\ 64&$\usym{2717}$&$\usym{2713}$\\
            P2I-MI~\cite{liu2024prediction} [2024]&StyleGAN&$\usym{2717}$&\labelicon&\bbicon&$\usym{2717}$&\rgbicon\ 64&$\usym{2717}$&$\usym{2713}$\\
            QE-MIA~\cite{xu2025query} [2025]&StyleGAN&$\usym{2717}$&\labelicon&\bothicon&$\usym{2717}$&\rgbicon\ 1024&$\usym{2713}$&$\usym{2717}$\\
            Shahreza~\etal~\cite{otroshi2023face} [2023]&StyleGAN&$\usym{2717}$&\embeddingicon&\bothicon&$\usym{2713}$&\rgbicon\ 1024&$\usym{2713}$&$\usym{2713}$\\
            Shahreza~\etal~\cite{shahreza2024template} [2024]\textcolor{red}{$^*$}&StyleGAN&$\usym{2717}$&\embeddingicon&\bothicon&$\usym{2713}$&\rgbicon\ 1024&$\usym{2713}$&$\usym{2713}$\\
            PPA~\cite{struppek2022plug} [2022]&StyleGAN&$\usym{2713}$&\labelicon&\wbicon&$\usym{2717}$&\rgbicon\ 1024&$\usym{2713}$&$\usym{2713}$\\
            IF-GMI~\cite{qiu2024closer} [2024]&StyleGAN&$\usym{2713}$&\labelicon&\wbicon&$\usym{2717}$&\rgbicon\ 224&$\usym{2717}$&$\usym{2713}$\\
            Dong~\etal~\cite{dong2023Reconstruct} [2023]&StyleGAN&$\usym{2713}$&\embeddingicon&\bbicon&$\usym{2713}$&\rgbicon\ 1024&$\usym{2713}$&$\usym{2717}$\\
            MAP$^2$V~\cite{zhang2024validating} [2024]\textcolor{red}{$^*$}&StyleGAN&$\usym{2713}$&\embeddingicon&\bothicon&$\usym{2713}$&\rgbicon\ 192&$\usym{2717}$&$\usym{2713}$\\
            PriDM~\cite{pang2025pridm} [2025]&DDPM&$\usym{2713}$&\imageicon&\bbicon&$\usym{2717}$&\rgbicon\ 256&$\usym{2713}$&$\usym{2717}$\\
            \midrule
            \textbf{DiffMI (Ours)}&DDPM&$\usym{2713}$&\embeddingicon&\bothicon&$\usym{2713}$&\rgbicon\ 256&$\usym{2713}$&$\usym{2713}$\\
            \bottomrule
        \end{tabular}}
        \begin{tablenotes}
            \item $^1$ \labelicon\ Label \quad \embeddingicon\ Embedding \quad \imageicon\ Partial raw facial image \quad\quad $^2$ \wbicon\ White-Box \quad \bbicon\ Black-Box \quad \bothicon\ Both \quad\quad $^3$ \grayscaleicon\ Grayscale \quad \rgbicon\ RGB
            \item \textcolor{red}{$^*$} Indicates benchmark methods used for empirical comparison, selected as representative exemplars of each strategy.
        \end{tablenotes}
    \end{threeparttable}
\end{table*}

However, recent model inversion attacks show that facial images can be reconstructed from embeddings alone, enabling threats such as identity spoofing in biometric authentication (\eg unlocking smartphones or accessing secure facilities), surveillance and re-identification in public datasets, and long-term identity theft. As illustrated in \cref{fig_mia}, even without raw photos, recovering a face from an embedding may allow attackers to forge IDs, impersonate users in video calls, or bypass bans on reusing facial data~\cite{boulkenafet2017oulu,tan2021many,wang2021similarity,wang2024multi}. Understanding these inversion risks is thus critical for assessing the real-world security of embedding-based systems~\cite{zhang2024validating}.

While attacks have demonstrated the feasibility, they continue to face several \textbf{key challenges}.

\emph{\one Training-Dependent:}
Most prior attacks are training-dependent, as listed in \cref{tab_relatedworks}, requiring the training or fine-tuning of a target-specific generator. Such strategy incurs substantial computational overhead while offering limited generalizability~\cite{fredrikson2015model,mai2018reconstruction,shahreza2024vulnerability,zhang2023analysis,nguyen2023label,zhang2020secret,khosravy2022model,yuan2023pseudo,wu2025label,duong2020vec2face,kansy2023controllable,liu2024unstoppable,otroshi2023face,shahreza2024template}. Model inversion typically relies on a generative prior. In this work, \emph{training-free} means that a single fixed pretrained generator~\cite{karras2020analyzing,meng2022sdedit} is reused across target models without target-specific training or fine-tuning~\cite{zhang2024validating,pang2025pridm}, in contrast to \emph{training-dependent} approaches that require a separate generator per target~\cite{shahreza2024vulnerability,nguyen2023label,liu2024unstoppable}.

Moreover, many training-dependent attacks target closed-set identity classification, where adding or removing identities requires retraining the recognition model. These attacks rely on task-specific training over a fixed set of identities (\ie seen models and seen identities)~\cite{fredrikson2015model,zhang2023analysis,nguyen2023label,zhang2020secret,khosravy2022model,yuan2023pseudo,wu2025label,liu2024unstoppable,struppek2022plug,qiu2024closer}. In contrast, modern face recognition systems are embedding-based and open-set by design, supporting generalization to unseen identities and dynamic user enrollment without retraining. As a result, closed-set attacks are poorly aligned with real-world deployment scenarios.

Additionally, high-fidelity reconstruction remains challenging for training-dependent methods due to their substantial training overhead~\cite{fredrikson2015model,mai2018reconstruction,shahreza2024vulnerability,zhang2023analysis,nguyen2023label,zhang2020secret,khosravy2022model,yuan2023pseudo,wu2025label,duong2020vec2face,kansy2023controllable,liu2024unstoppable}. Effective identity recovery requires visually rich, high-resolution outputs (ideally with larger image size and full headshot-style, \aka selfie, reconstructions)~\cite{dong2023Reconstruct,otroshi2023face,shahreza2024template,struppek2022plug}. As shown in \cref{tab_relatedworks}, training-free methods more easily achieve this, while training-dependent ones often produce only low-resolution (\eg $64{\times}64$) or grayscale facial crops, insufficient for reliable recognition.

\emph{\two Limited Identity Controllability:}  
Even training-free, GAN-based attacks suffer from limited identity controllability~\cite{struppek2022plug,qiu2024closer,dong2023Reconstruct,zhang2024validating}. Despite the prevalence of diffusion models in modern generative systems~\cite{rombach2022high}, training-free inversion attacks still largely rely on GANs~\cite{karras2020analyzing}, which are typically constrained by a truncation radius to suppress artifacts, limiting fine-grained identity control.

\emph{\three Adversarial Overfitting:}  
Naively adapting GAN-style inversion to diffusion causes severe artifacts and overfitting. A simple training-free approach applies adversarial attacks directly in the diffusion latent space, but such updates often push latents off the Gaussian prior, causing the reverse denoising process to amplify off-manifold signals into halos, patchy textures, and model-specific overfitting. These artifacts may even mislead the target model into falsely confirming reconstruction success. Consequently, prior diffusion-based attacks rely on \emph{conditional} generation, which requires target-specific training, days of computation, and yields only low-resolution results~\cite{kansy2023controllable,liu2024unstoppable}.

\textbf{Contributions:} Addressing these limitations requires a training-free approach that can guide unconditional diffusion models toward identity-consistent reconstructions while suppressing overfitting-induced artifacts. Our contributions are summarized as follows:
\begin{itemize}[leftmargin=*]
\item \textbf{The first diffusion-driven, training-free model inversion.}  
We introduce DiffMI, the first diffusion-driven, training-free inversion attack on embedding-based face recognition. It uses a fixed, pretrained diffusion model to generate full headshot-style images directly from embeddings, without target-specific generator training.

\item \textbf{A principled identity-preserving inversion mechanism.}
We identify three key factors for success: robust latent initialization, fine-grained adversarial manipulation, and artifact suppression. Our pipeline integrates automated robust latent code selection, a ranked adversarial refinement strategy, and a statistically grounded confidence-aware objective. These components enable precise latent-space updates with principled stopping to avoid overfitting, supporting attacks on unseen identities and models.

\item \textbf{State-of-the-art performance.}  
On two datasets~\cite{Lee2020mask,hua2008labeled} and four recognition models~\cite{schroff2015facenet,deng2019arcface,ji2022privacy,mi2023privacy}, DiffMI exceeds the best prior training-free baseline~\cite{zhang2024validating} by +8.97\% in identity match accuracy. It also compromises inversion-resilient systems~\cite{ji2022privacy,mi2023privacy}, achieving 84.42\%--92.87\% success rates, revealing critical vulnerabilities in current privacy defenses. We further conduct \emph{user studies} in which participants assess whether reconstructions match target identities.
\end{itemize}

\begin{figure}[!t]
    \centering
    \includegraphics[width=3in]{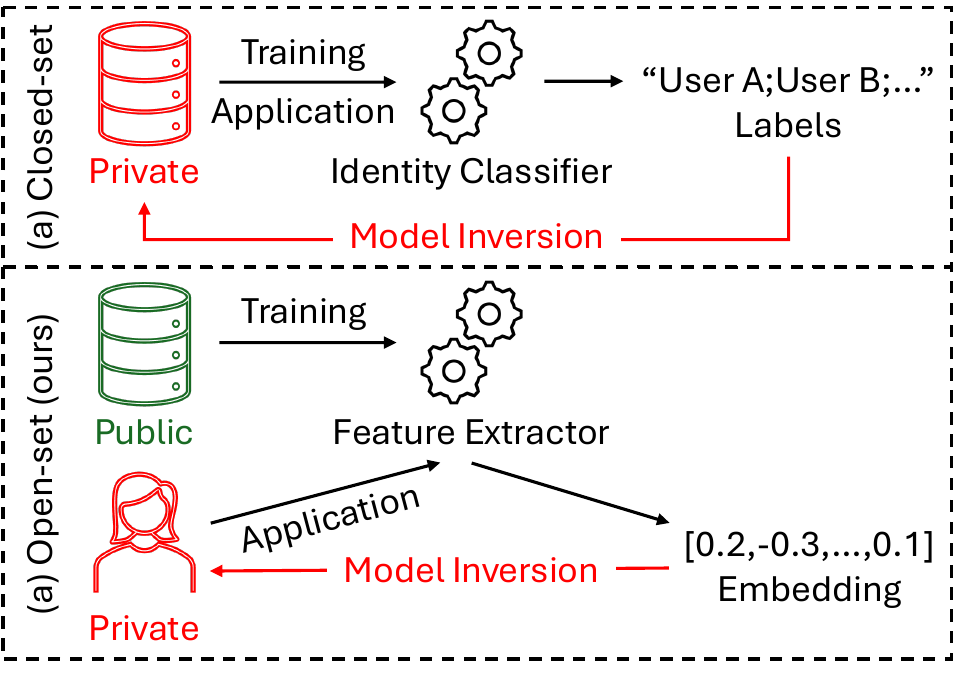}
    \caption{Comparison of label-based closed-set and embedding-based open-set model inversion, highlighting differences in private data, model usage, and inversion inputs.}
    \label{fig_close_vs_open}
\end{figure}

\section{Preliminary}

\subsection{Label-Based Closed-Set vs. Embedding-Based Open-Set Model Inversion}

As illustrated in \cref{fig_close_vs_open} and summarized in \cref{tab_relatedworks}, model inversion attacks can be categorized into \emph{label-based closed-set} and \emph{embedding-based open-set} paradigms, which differ in private data, model usage, and inversion objectives.

In closed-set classification, the private data is the model’s training dataset, and inversion aims to reconstruct samples of those fixed identities from class labels or scores. In contrast, under the embedding-based open-set setting, the private data corresponds to enrolled user faces whose embeddings are stored for verification. The backbone is typically trained on public datasets, and attacked identities are not assumed to belong to the training set.

Closed-set models perform identity classification over predefined classes, whereas embedding-based models act as feature extractors mapping faces into a continuous space for similarity comparison. Accordingly, the inversion input shifts from labels to embeddings, and the objective becomes synthesizing an identity-consistent image rather than reconstructing a specific training sample.

\subsection{Why Embedding-Based Open-Set Recognition?}

Embedding-based recognition reflects modern practice and realistic privacy risks. Models such as FaceNet~\cite{schroff2015facenet} and ArcFace~\cite{deng2019arcface} enable scalable open-set recognition by comparing embeddings rather than classifying fixed identities. Commercial systems (e.g., Microsoft Azure~\cite{MicrosoftAzureFaceAPI}, Amazon Rekognition~\cite{AmazonRekognition}, Google Cloud Vision~\cite{GoogleCloudVisionAPI}) similarly store embeddings as persistent identity representations. Studying embedding-based inversion therefore aligns with real-world deployments, where embeddings—not training images—constitute the primary sensitive artifact.


\section{Diffusion-Driven Model Inversion (DiffMI)}
\cref{threatmodel} introduces the threat model. \cref{overview} outlines the objectives and overall framework (\cref{fig_DiffMI}). \cref{problem_loss} formalizes the problem and defines the objective function. The three core algorithmic components, including latent code initialization, selection, and manipulation, are detailed in \cref{latentgeneration,TopN,manipulaiton}. Notations are summarized in the supplementary materials.

\subsection{Threat Model}
\label{threatmodel}
DiffMI functions as an attacker that assesses the privacy vulnerabilities of a face recognition model. Specifically, it examines whether a facial image reconstructed solely from a feature embedding can visually resemble the identity represented by that embedding.

\subsubsection{Attack Knowledge}
The attacker's level of access determines the threat model:
\begin{itemize}
    \item \emph{\wbicon \quad White-box:} Access to the target embedding and the feature extractor, including gradient information.
    \item \emph{\bbicon \quad Black-box:} Access to the target embedding and query-based access to the model for similarity computation, without access to internal parameters or gradients.
\end{itemize}
In both settings, the attacker has no access to identity labels, raw images, or original training data. The only private information assumed to be available is the target embedding.

\subsubsection{Attack Objective} Reconstructs a face that enables recognition of the target identity, assessed via two criteria:
\begin{itemize}
    \item The reconstruction is visually perceived as depicting the same person as the target.
    \item The reconstruction enables identification of the target among a set of identities.
\end{itemize}

While visual similarity is inherently subjective and best assessed via user studies, we complement human evaluation with automated metrics using face recognition. Because adversarial optimization may introduce artifacts exploiting model-specific weaknesses, we assess DiffMI’s effectiveness through \emph{cross-model robustness}, a widely adopted privacy risk metric measuring whether reconstructed faces achieve transferable identity alignment across \emph{non-target models}~\cite{zhang2024validating,shahreza2024vulnerability,otroshi2023face,shahreza2024template,dong2023Reconstruct}. Formally, the attack is defined by two objectives:
\begin{itemize}[leftmargin=*]
    \item \emph{During optimization:} Maximize the similarity between embeddings (\aka~embedding similarity) of the reconstruction and the target until a predefined confidence threshold is reached.
    \item \emph{During evaluation:} Ensure the output is recognized as the target identity by non-target models, not only against the specific image used for inversion but also against other real images of the same person. This ensures the reconstruction reflects the underlying identity rather than overfitting to a single reference image.
\end{itemize}

\begin{figure*}[!t]
    \centering
    \includegraphics[width=5.1in]{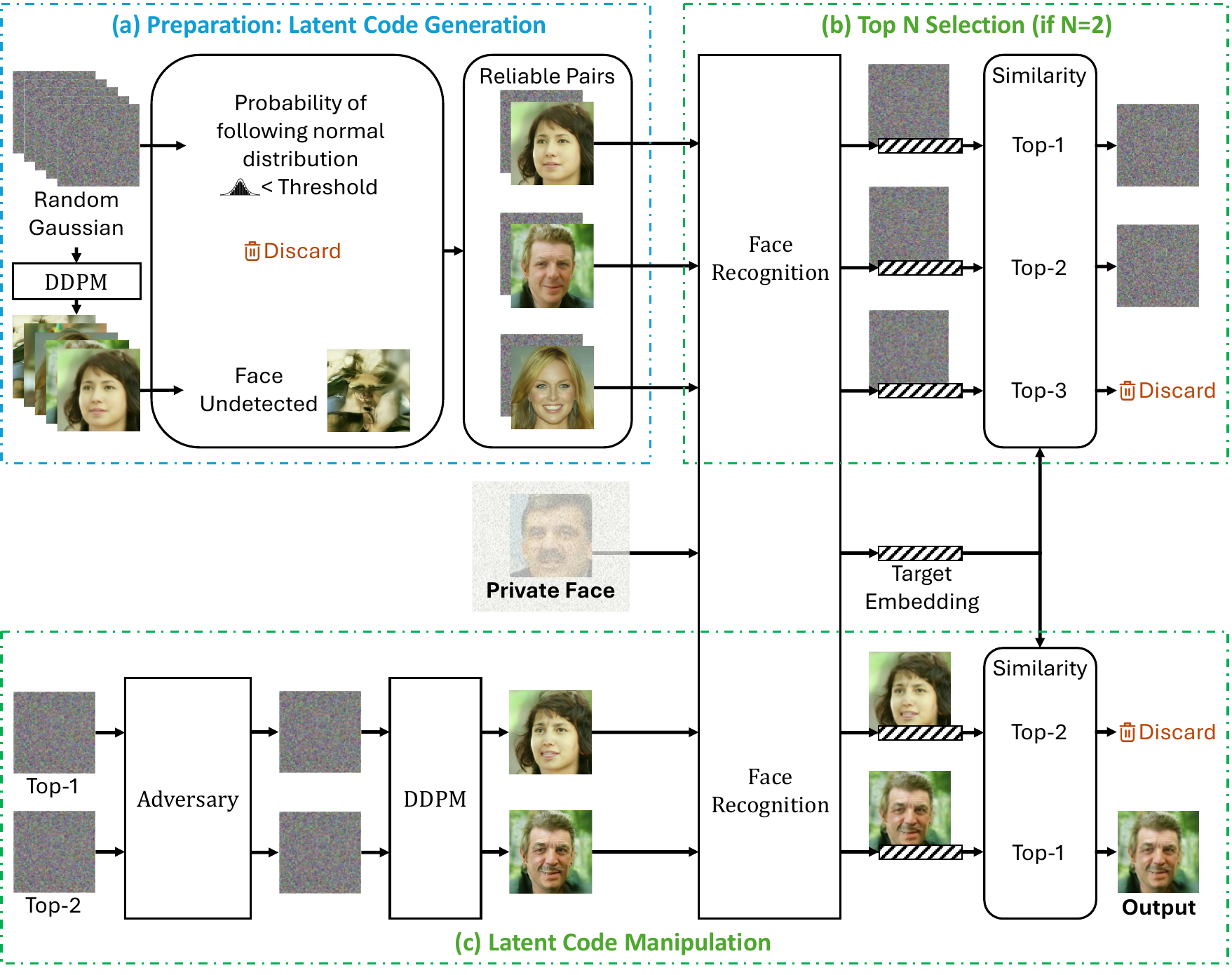}
    \caption{Framework of DiffMI, which reconstructs a facial image sharing the same identity as a private face solely from its embedding. The generator, a denoising diffusion model (DDPM)~\cite{meng2022sdedit}, is pretrained independently, without prior knowledge of the private face or the target model. In Step~(a), a set of highly robust latent codes is generated once and reused for any target. In Step~(b), $N$ codes whose reconstructions have the highest similarity to the target embedding are selected as initialization for Step~(c). In Step~(c), these codes undergo adversarial refinement to progressively align the reconstructions with the target embedding. Notably, a higher initial rank does not guarantee superior final reconstruction after refinement.}
    \label{fig_DiffMI}
\end{figure*}

\subsection{Overview}
\label{overview}
DiffMI is a training-free framework for assessing the privacy risks of embedding-based face recognition models. A model is deemed vulnerable if DiffMI can reconstruct a facial image resembling the target identity from its feature embedding alone. The goal is to recover the identity associated with the embedding, not the original input image. Built on DDPM~\cite{meng2022sdedit}, a pretrained unconditional face generator, DiffMI generalizes to unseen target models and identities.

The key challenge is guiding unconditional diffusion toward identity-consistent outputs while suppressing adversarial artifacts. We address this with a \textbf{confidence-aware objective} that enables early stopping to prevent overfitting, coupled with the following three-step procedure (\cref{fig_DiffMI}):

\emph{Step (a) – Robust Latent Code Initialization (Independent Preparation).}  
A pool of robust latent codes is prepared for reuse in subsequent attacks. Robustness is enforced via a dual-selection strategy: D'Agostino's $K^2$ test~\cite{d1970transformation,d1973tests,d1990suggestion} ensures Gaussian normality of the latent distribution, and MTCNN~\cite{zhang2016joint} verifies the presence of facial features in the initial generations. These filters improve initialization quality and resistance to optimization-induced distortion.

\emph{Step (b) – Top $N$ Latent Code Selection.}  
From the filtered pool, the top $N$ codes are selected by embedding similarity between their generated images and the target. This provides stronger identity alignment before manipulation, even if the images are not yet close matches.

\emph{Step (c) – Ranked, Fine-Grained Latent Code Manipulation.}  
The selected codes undergo fine-grained adversarial optimization to maximize confidence-aware embedding similarity with the target. A confidence threshold enables early stopping once the similarity score on the target model reaches a statistically grounded level, indicating sufficient refinement for cross-model robustness. A ranked adversary strategy, guided by similarity priorities and the confidence-aware objective, improves computational efficiency and mitigates overfitting-induced artifacts.

\subsection{Problem Definition and Objective Function}
\label{problem_loss}
Embedding-based face recognition models map input images $x$ to $d$-dimensional embeddings $z = F(x) \in \mathbb{R}^d$, where $F(\cdot)$ denotes the embedding function, representing the face recognition model. Two images are classified as the same identity if the cosine similarity of their embeddings satisfies:
\(
    S(z_1, z_2) = \frac{z_1 \cdot z_2}{\|z_1\| \|z_2\|} \geq \tau_F,
\)
where $\tau_F$ is the decision threshold optimized for the minimum equal error rate (EER).

Given a target face $x^\mathrm{tgt}$ and target model $F(\cdot)$, DiffMI seeks to reconstruct an image $\hat{x}$ sharing the identity of $x^\mathrm{tgt}$, using only its embedding $z^\mathrm{tgt} = F(x^\mathrm{tgt})$. The objective is identity preservation rather than pixel-level fidelity:
\begin{equation}
    S\left(F(\hat{x}), F(x^\mathrm{tgt})\right) \geq \tau_F.
    \label{eq_problem_F}
\end{equation}

The reconstruction $\hat{x}$ is produced by a pretrained DDPM~\cite{meng2022sdedit}, modeled as a generative function $G(\cdot)$ applied to a latent code $x_G$ drawn from a standard Gaussian distribution:
\begin{equation}
    \hat{x} = G(x_G), \quad x_G \sim \mathcal{N}(0, I).
\end{equation}
Substituting into \cref{eq_problem_F} gives:
\begin{equation}
    S\left(F(G(x_G)), F(x^\mathrm{tgt})\right) \geq \tau_F.
    \label{eq_problem_G}
\end{equation}

Since a randomly sampled $x_G$ rarely satisfies \cref{eq_problem_G}, we introduce an adversarial perturbation $\delta$, constrained by an $L_p$-norm bound $\epsilon$, to optimize $x_G$:
\begin{equation}
    x_G' = x_G + \delta, \quad \text{s.t.} \quad \|\delta\|_p \leq \epsilon,
    \label{eq_xadv}
\end{equation}
yielding the refined objective:
\begin{equation}
    S\left(F(G(x_G + \delta)), F(x^\mathrm{tgt})\right) \geq \tau_F.
    \label{eq_problem}
\end{equation}
Therefore, the objective function $\mathcal{L}$ is defined via \cref{eq_problem}:
\begin{equation}
    \mathcal{L} = S\left(F(G(x_G + \delta)), F(x^\mathrm{tgt})\right).
    \label{eq_loss}
\end{equation}
The goal is to iteratively update $\delta$ to maximize $\mathcal{L}$:
\begin{equation}
    \delta = \underset{\|\delta\|_p \leq \epsilon}{\arg\max}\ \mathcal{L}.
    \label{eq_obj}
\end{equation}

\begin{figure}[!b]
    \centering
    \includegraphics[width=2.8in]{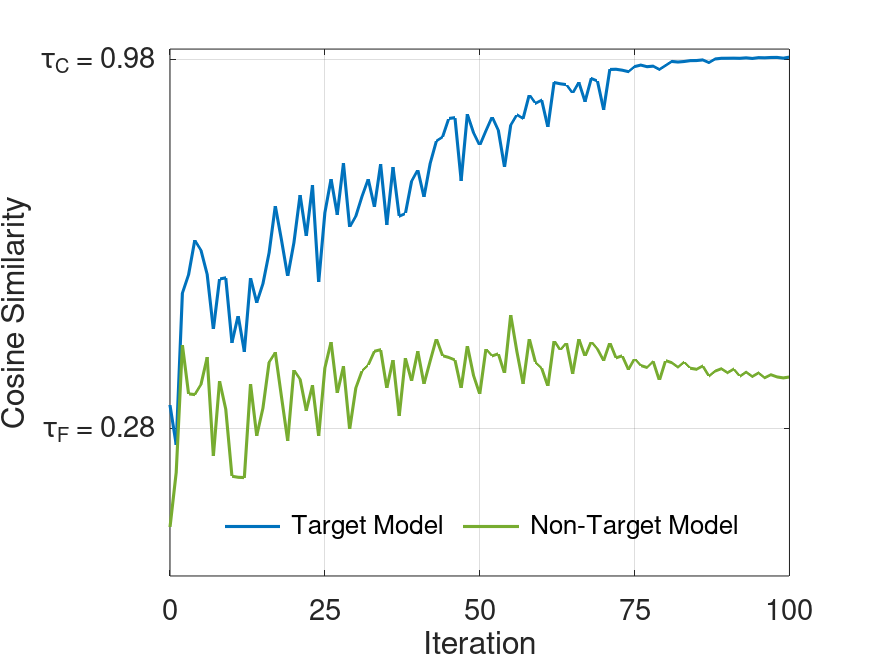}
    \caption{Loss convergence on the target model and cross-model robustness on the non-target model. To achieve stable cross-model robustness, the optimization must push $\mathcal{L}$ well beyond the target model’s decision threshold $\tau_F$. However, once $\mathcal{L} > \tau_C$, further updates yield diminishing improvements in similarity while degrading cross-model robustness.}
    \label{fig_tauC}
\end{figure}

\textbf{Confidence-Aware Objective}
As shown in \cref{fig_tauC}, stable cross-model robustness requires the similarity score $\mathcal{L}$ to substantially exceed the target model’s decision threshold $\tau_F$. Although the objective in \cref{eq_obj} steadily increases $\mathcal{L}$ on the target model, further updates beyond a certain point (\eg $\mathcal{L} > 0.98$) offer negligible gains while degrading generalization. This indicates that blindly maximizing $\mathcal{L}$ adds unnecessary computation and risks overfitting to the target model.

To address this, we propose a \emph{confidence-aware objective} with an early stopping criterion. A confidence threshold $\tau_C$ is set, and optimization terminates once $\mathcal{L} \geq \tau_C$. This threshold is a statistically grounded upper bound on the maximum similarity observed between real images of the same identity under the target model:
\begin{equation}
    \tau_C = \max\ S\left(F(x^i), F(x^{j \neq i})\right), \quad x^i, x^j \in \mathcal{X}_{real}.
    \label{eq_tauC}
\end{equation}

The optimal $\tau_C$ is computed once per target model and does not require per-dataset or per-identity tuning. As shown in \cref{ablation_margin}, DiffMI is not highly sensitive to its exact value. In practice, $\tau_C$ should avoid being too low or too close to $1.0$, and \cref{eq_tauC} serves as a statistical guideline rather than a strict requirement.

\begin{figure}[!t]
    \centering
    \includegraphics[width=3in]{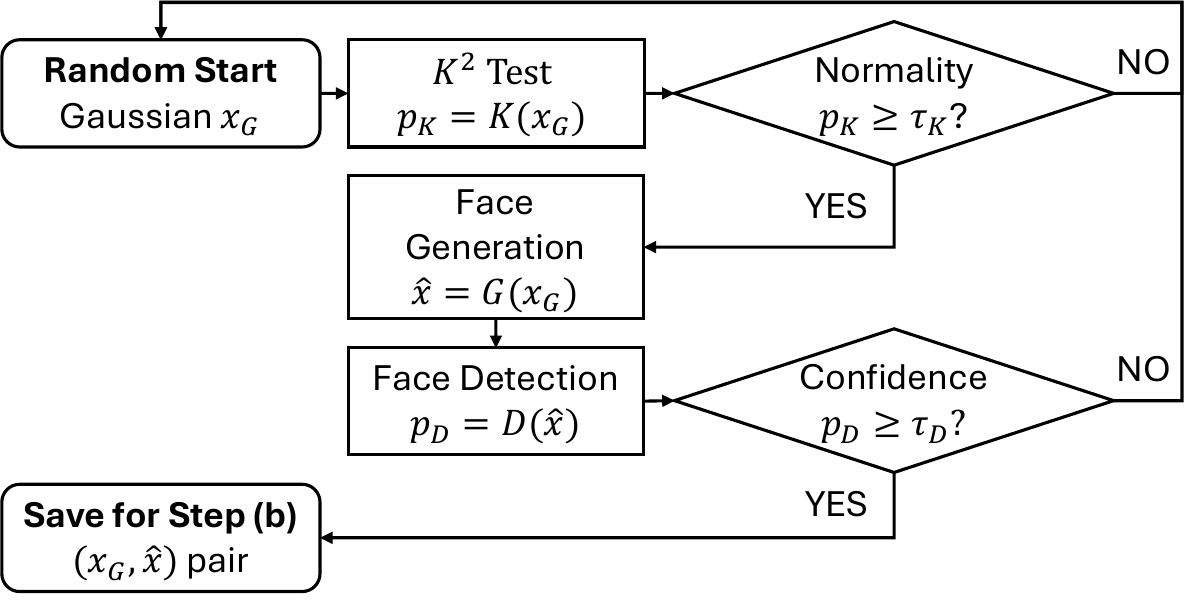}
    \caption{Two-stage latent code generation. First, $K^2$ test $K(\cdot)$~\cite{d1970transformation, d1973tests, d1990suggestion} filters each sampled latent code $x_G$ based on Gaussian normality, retaining those with $p_K \geq \tau_K$. Then, MTCNN $D(\cdot)$~\cite{zhang2016joint} verifies whether the corresponding image $\hat{x} = G(x_G)$ contains a detectable face, accepting codes with detection confidence $p_D \geq \tau_D$.}
    \label{fig_LC_generate}
\end{figure}

\subsection{Step (a) – Latent Code Initialization}
\label{latentgeneration}

We adopt a two-stage strategy to generate robust latent codes for initialization (\cref{fig_LC_generate}), ensuring both statistical validity and visual plausibility. This involves filtering with D'Agostino's $K^2$ test~\cite{d1970transformation,d1973tests,d1990suggestion} for Gaussianity and verifying facial presence via MTCNN~\cite{zhang2016joint}. Step (a) is performed once to build a reusable pool of high-quality latent codes applicable across different targets. Enforcing both statistical and perceptual constraints ensures robust initialization, reducing susceptibility to distortion during later adversarial optimization.

The process proceeds sequentially: sampled latent codes are first screened via the $K^2$ test, a lightweight check discarding those deviating from the Gaussian prior. Remaining codes are then denoised into images $\hat{x} = G(x_G)$ and passed through MTCNN for face detection, a more computationally expensive step. This ordering improves efficiency, as statistically valid codes are likelier to yield recognizable faces.

\subsubsection{Normality Test via D'Agostino's K-Square Statistic}
\label{ktest}
Reconstructing facial images with DDPM requires latent codes of size $3 \times 256 \times 256$ to follow a normal distribution (\ie Gaussian normality)~\cite{meng2022sdedit}. The normality of randomly sampled codes varies and generally correlates with reconstruction quality. However, even highly normal codes will be pushed off this manifold after adversarial manipulation, with deviations amplified during denoising and leading to visual artifacts. Evidence of normality variance and its deterioration after manipulation is shown in \cref{fig_normality,fig_normality_gap} in ablation.

Although even highly normal initial codes will inevitably drift off the diffusion prior manifold during adversarial refinement, starting from latent codes that adhere more closely to a Gaussian distribution reduces the magnitude of distortion. Therefore, we apply D'Agostino's $K^2$ test~\cite{d1970transformation,d1973tests,d1990suggestion} to select latent codes with stronger initial normality, which improves baseline image quality and preserves fidelity throughout optimization. While this cannot fully prevent deviation, it effectively mitigates artifact accumulation. Given a latent code $x_G$, the $K^2$ test function $K(\cdot)$ computes a $p$-value based on skewness and kurtosis:
\begin{equation}
    p_K = K(x_G).
\end{equation}
We retain codes satisfying the normality threshold $\tau_K$:
\begin{equation}
    x_G \text{ is selected if } p_K \geq \tau_K.
\end{equation}
Since latent sampling is inexpensive, a strict threshold can be enforced without limiting code availability. This filtering step ensures selected codes adhere closely to Gaussian properties, improving robustness to subsequent adversarial perturbations.

\subsubsection{Face Detection via MTCNN}
\label{face_detection}
Latent codes satisfying Gaussian normality may still fail to produce recognizable faces (see examples in \cref{fig_highnormality_fail} in ablation). We use MTCNN~\cite{zhang2016joint}, a deep learning-based face detector, to ensure the presence of facial features.

Given a latent code $x_G$, the DDPM generator $G(\cdot)$ produces the corresponding image:
\begin{equation}
    \hat{x} = G(x_G).
\end{equation}
We then apply the face detector $D(\cdot)$, which returns a confidence score $p_D$ indicating the likelihood of a detected face:
\begin{equation}
    p_D = D(\hat{x}).
\end{equation}
Latent codes are accepted if their detection confidence exceeds a predefined threshold $\tau_D$:
\begin{equation}
    (x_G, \hat{x}) \text{ is selected if } p_D \geq \tau_D.
\end{equation}

Since most latent codes passing $K^2$ test already yield plausible facial structures, a relatively high $\tau_D$ can be used without significantly reducing the acceptance rate. To support later stages, both $x_G$ and $\hat{x}$ are cached for efficient reuse.

\begin{figure}[!t]
    \centering
    \includegraphics[width=\linewidth]{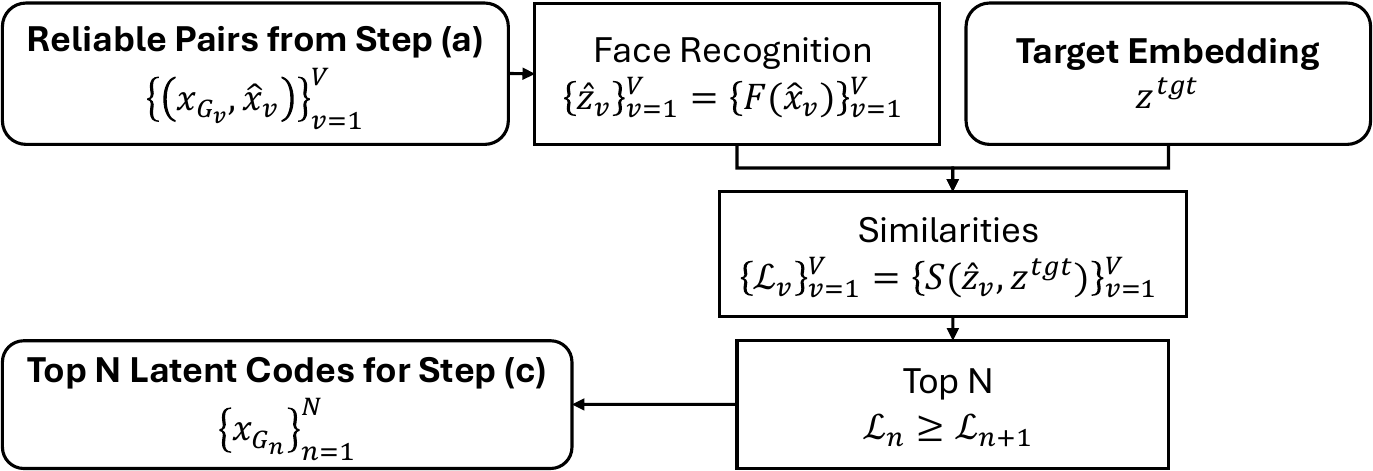}
    \caption{Selection of the top $N$ latent codes based on embedding similarity to the target identity.}
    \label{fig_TopN}
\end{figure}

\begin{figure*}[!t]
    \centering
    \includegraphics[width=6in]{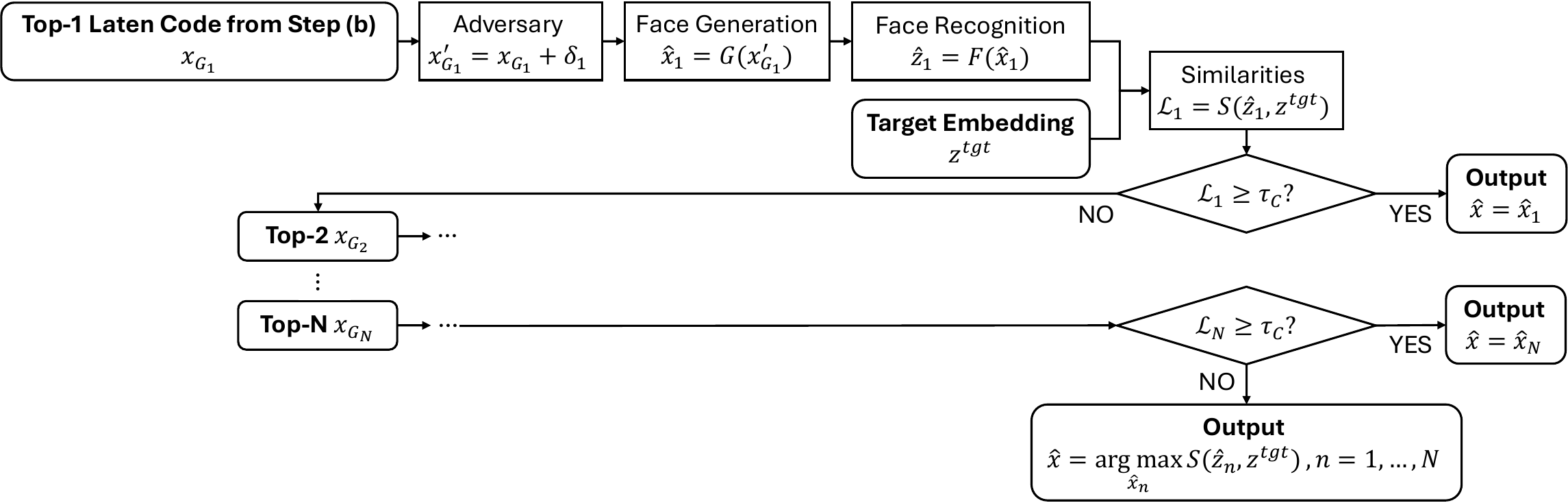}
    \caption{Ranked Adversary algorithm for latent code manipulation. The top $N$ latent codes, ranked in Step (b), are sequentially optimized via adversarial manipulation to maximize the objective in \cref{eq_loss}. The process terminates early if any code satisfies $\mathcal{L}_n \geq \tau_C$ for $n = 1, \dots, N$. If none meet the confidence threshold $\tau_C$, the code $x'_{G_n}$ with the highest $\mathcal{L}_n$ is selected, and its reconstruction is returned as the final facial image.}
    \label{fig_manipulation}
\end{figure*}

\subsection{Step (b) – Latent Code Selection}
\label{TopN}
Step (a) yields $V$ robust latent codes and their reconstructions, $\{(x_{G_v}, \hat{x}_v)\}_{v=1}^V$. Selecting the single code with the highest initial similarity to the target embedding minimizes computation but may underperform after adversarial optimization (see examples in \cref{fig_TopN_fail} in ablation). To address this, DiffMI selects the top $N$ candidates ($N \leq V$) ranked by identity similarity (\cref{fig_TopN}). Larger $N$ generally improves inversion success at the cost of higher computation.

Specifically, each reconstruction $\hat{x}_v = G(x_{G_v})$, precomputed in Step (a), is passed through the target model $F(\cdot)$ to obtain an embedding. Cosine similarity (\cref{eq_loss}) is then computed between this embedding and the target embedding $z^\mathrm{tgt}$. The top $N$ latent codes with the highest similarity scores are retained as initialization points for Step (c).

In the black-box setting (\cref{threatmodel}), Step (b) requires $V$ model queries, one per candidate:
\begin{equation}
    Q_{TopN} = V,
    \label{eq_q_TopN}
\end{equation}
where $V$ is the number of latent code candidates. Notably, the cost in Step~(b) is $V$ rather than $N$, as it requires computing similarity scores between all $V$ latent codes and the target embedding in order to select the top $N$ candidates.

\subsection{Step (c) – Latent Code Manipulation}
\label{manipulaiton}
Randomly sampled latent codes, even those close to the target embedding, typically require refinement to satisfy the identity-matching objective. Prior methods either refine only the top-1 code or fuse the top $N$ candidates~\cite{zhang2024validating}, but fusion often reduces identity similarity and shares the same failure modes as single-code strategies, where the best initial match does not guarantee the best final result.

To address these limitations, we propose the \emph{Ranked Adversary} strategy: a ranking-based refinement approach that sequentially optimizes the top $N$ latent codes, prioritizing those with higher initial similarity. This avoids fusion, preserves candidate diversity, and enables early stopping via the confidence-aware objective.

\subsubsection{Ranked Adversary}
\label{ranked_adv}
The Ranked Adversary algorithm (\cref{fig_manipulation}) sequentially refines the top $N$ latent codes obtained from Step~(b), starting with the highest-ranked code $x_{G_1}$. Each code is updated adversarially to maximize identity similarity, as defined in \cref{eq_loss}.

The process stops early if any refined code satisfies the confidence threshold, \ie $\mathcal{L}_n \geq \tau_C$ for some $n \in \{1, \dots, N\}$, returning $\hat{x}_n = G(x_{G_n} + \delta_n)$ as the final reconstruction. If no code reaches the threshold within the maximum number of iterations $t_{\max}$, the code with the highest final similarity is selected, and its reconstruction is returned as the output.

\subsubsection{Fine-Grained Adversarial Learning}
Effective manipulation of unconditional diffusion generation does not require novel adversarial architectures, but rather a fine-grained optimization strategy. As defined by Wang~\etal~\cite{wang2025greedypixel}, fine-grained adversarial attacks operate at a localized, precision-controlled level, enabling gradual refinement in high-dimensional latent spaces. The Ranked Adversary employs APGD~\cite{croce2020reliable} for white-box and GreedyPixel~\cite{wang2025greedypixel} for black-box settings.

While alternative coarse-grained methods can be integrated into our manipulation pipeline, empirical results (\cref{fig_DiffMI_black}) show that only fine-grained techniques consistently produce high-quality reconstructions without significant visual artifacts. This behavior stems from the sensitivity of the diffusion model’s latent space: even small off-manifold perturbations can propagate through the denoising process and amplify into perceptible degradation. Unlike classifier-based models, where local smoothness may tolerate coarse updates, the iterative nature of diffusion models compounds small inconsistencies, making precision essential.

\subsubsection{Query Efficiency in the Black-Box Setting}
In the black-box setting, query cost arises from computing the loss value (\cref{eq_loss}), which requires querying the target model to obtain embeddings of reconstruction. Consequently, the total query cost is proportional to the number of adversarial iterations.

For a single latent code, the query cost is:
\begin{equation}
    Q_{Adv} \leq t_{\max},
\end{equation}
where $t_{\max}$ is the maximum number of adversarial iterations. Since the Ranked Adversary may process up to $N$ latent codes, the total query cost for the manipulation phase is bounded by:
\begin{equation}
    Q_{Adv} \leq N \times t_{\max}.
    \label{eq_q_adv}
\end{equation}

Including the $V$ queries for top-$N$ latent code selection in Step~(b) (\cref{eq_q_TopN}), the total query complexity of DiffMI is:
\begin{equation}
    Q = Q_{TopN} + Q_{Adv}.
    \label{eq_q}
\end{equation}

In practice, black-box attacks are constrained by a strict query budget $Q_{\max}$. To satisfy this constraint, the number of iterations per latent code must be bounded by:
\begin{equation}
    t_{\max} = \left\lfloor \frac{Q_{\max} - V}{N} \right\rfloor,
    \label{eq_tmax}
\end{equation}
where $\lfloor \cdot \rfloor$ denotes the floor function, and $V$ is the number of latent codes evaluated in Step (b). This formulation enables precise control over query consumption while maintaining optimal attack effectiveness within the given budget.

\begin{table*}[!t]
    \footnotesize
    \centering
    \begin{threeparttable}
        \caption{Diversity of Model Architectures and Datasets.}
        \label{tab_diversity}
        \setlength{\tabcolsep}{5.7mm}{\begin{tabular}{ccccc}
            \toprule
            Model&Architecture&Training Dataset&Attack Dataset&Use\\
            \midrule
            FaceNet~\cite{schroff2015facenet}&InceptionResNetV1&VGGFace2~\cite{cao2018vggface2}&\multirow{7.5}{*}{\begin{tabular}{c}LFW~\cite{hua2008labeled} \textbf{(Major)}\\CelebA-HQ~\cite{Lee2020mask}\end{tabular}}&\multirow{4}{*}{Face Recognition Models}\\
            ArcFace~\cite{deng2019arcface}&IResNet100&MS1MV2~\cite{guo2016ms}&&\\
            DCTDP~\cite{ji2022privacy}&ResNet50&VGGFace2~\cite{cao2018vggface2}&&\\
            PartialFace~\cite{mi2023privacy}&IResNet50&MS1MV2~\cite{guo2016ms}&&\\
            \cmidrule{1-3}
            \cmidrule{5-5}
            DSCasConv~\cite{shahreza2024vulnerability}&DeconvNet&FFHQ~\cite{karras2019style}&&\multirow{3}{*}{Face Generators}\\
            StyleGAN~\cite{karras2020analyzing}&StyleGAN&FFHQ~\cite{karras2019style}&&\\
            DDPM~\cite{meng2022sdedit}&UNet&CelebA-HQ~\cite{Lee2020mask}&&\\
            \bottomrule
        \end{tabular}}
    \end{threeparttable}
\end{table*}

\begin{table}[!t]
   \footnotesize
    \centering
    \begin{threeparttable}
        \caption{Configurations of target models and DiffMI.}
        \label{tab_settings}
        \setlength{\tabcolsep}{0.4mm}{\begin{tabular}{cc|c|c|c|c|c|c|c|c}
            \toprule
            \multicolumn{2}{c|}{Face Recognition}&\multicolumn{3}{c|}{Latent Code Generation$^*$}&\multicolumn{5}{c}{Latent Code Manipulation}\\
            $F(\cdot)$&$\tau_F$&Volume $V$&$\tau_K$&$\tau_D$&Top $N$&$t_{max}$&$\tau_C$&Norm&$\epsilon$\\
            \midrule
            FaceNet \cite{schroff2015facenet}&0.40&\multirow{4}{*}{1,000}&\multirow{4}{*}{0.999}&\multirow{4}{*}{0.999}&\multirow{4}{*}{3}&\multirow{4}{*}{100}&0.99&\multirow{4}{*}{$L_2$}&25\\
            ArcFace \cite{deng2019arcface}&0.23&&&&&&0.99&&35\\
            DCTDP \cite{ji2022privacy}&0.26&&&&&&0.98&&35\\
            PartialFace \cite{mi2023privacy}&0.28&&&&&&0.98&&35\\
            \bottomrule
        \end{tabular}}
        \begin{tablenotes}
            \item $\tau_F$: Identity similarity threshold, set to minimize EER; $\tau_K$: Gaussian normality threshold; $\tau_D$: Detection confidence threshold; $\tau_C$: Optimization confidence threshold, computed in \cref{eq_tauC}. $^*$All experiments use a fixed latent code pool, except for the latent-generation ablation (\cref{ablation_latent}).
        \end{tablenotes}
    \end{threeparttable}
\end{table}

\section{Experimental Settings}
\label{settings}
Model architectures and datasets are summarized in \cref{tab_diversity}. White-box configurations are summarized in \cref{tab_settings}, and black-box settings are illustrated in \cref{fig_DiffMI_black}. All experiments are conducted on an NVIDIA A100 GPU.

\textbf{A. Face Recognition Models.} 
We evaluate inversion attacks on two widely used face recognition models: FaceNet~\cite{schroff2015facenet} and ArcFace~\cite{deng2019arcface}; as well as two inversion-resilient models: DCTDP~\cite{ji2022privacy} and PartialFace~\cite{mi2023privacy}, both designed to mitigate inversion. Decision thresholds are set at the minimum EER for each model, as detailed in \cref{tab_settings}.

\begin{figure}[!t]
    \centering
    \includegraphics[width=2.6in]{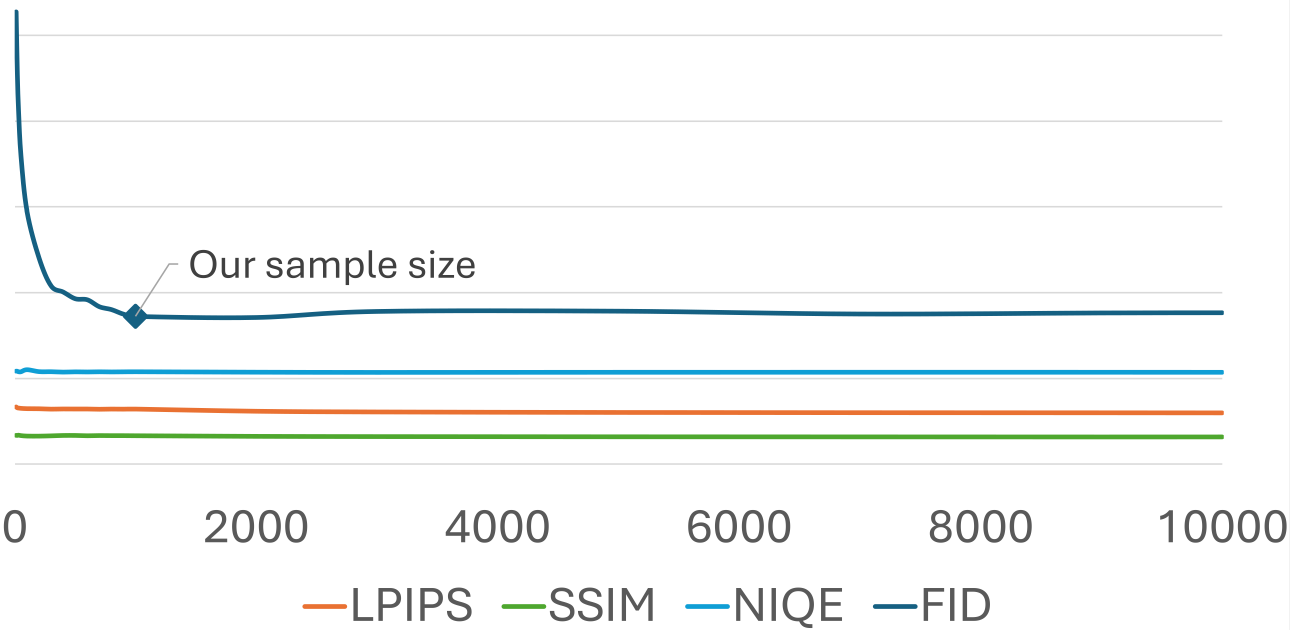}
    \caption{Statistical analysis of sample size for image quality evaluation. All metrics stabilize as the sample size increases, with negligible variation beyond approximately 1,000 samples.}
    \label{fig_sample_size}
\end{figure}

\textbf{B. Datasets.} 
We conduct evaluations on the LFW~\cite{hua2008labeled} and CelebA-HQ~\cite{Lee2020mask} datasets. Each dataset is subsampled to 1,000 images (10 images from 100 identities). Since DiffMI is training-free, evaluation variance is largely independent of dataset size; thus, 1,000-image subsets suffice for fair comparison. We further verify that all image quality metrics are stable at this scale (see \cref{fig_sample_size}). Datasets lacking identity labels, such as FFHQ~\cite{karras2019style}, are excluded.

\textbf{C. Baselines.} 
We compare DiffMI with three model inversion attacks spanning key design paradigms: a training-free GAN-based method (MAP$^2$V)~\cite{zhang2024validating}, a training-dependent DeconvNet-based method~\cite{shahreza2024template}, and a training-dependent GAN-based method~\cite{shahreza2024vulnerability}. These baselines capture the diversity of prior approaches across training-free and training-dependent settings. Closed-set attacks requiring target-specific training on a limited set of predefined identities are excluded.

\begin{figure}[!t]
    \centering
    \includegraphics[width=\linewidth]{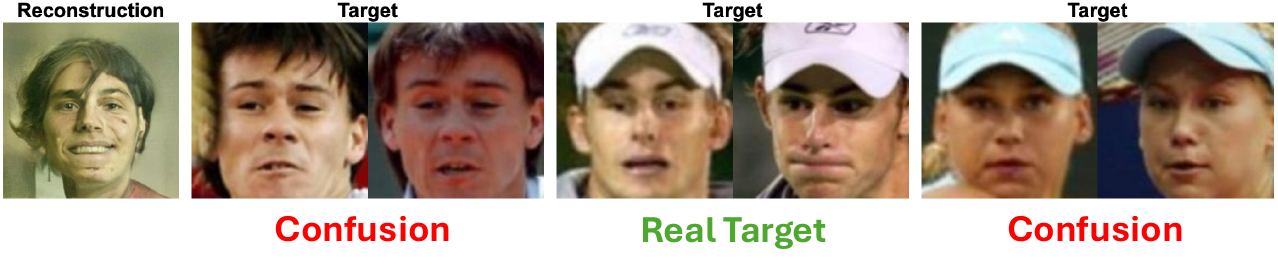}
    \caption{Example of difficult identity selection in the user study. Each candidate is represented by two photos; three identities are shown in randomized order. Several participants noted that the first two male candidates appeared too similar.}
    \label{fig_questionnaire}
\end{figure}

\textbf{D. User Study.} 
Following the objectives in \cref{threatmodel}, we conducted a user study with ten participants
tasked to: \one judge whether a reconstructed image depicts the same individual as the target (a straightforward matching task relevant to identity authentication); and \two identify the target individual from a candidate pool based on the reconstruction (mimicking witness identification in forensic scenarios). Each session comprised 60 images/identities from LFW and CelebA-HQ (MAP$^2$V evaluated with only 30). To increase difficulty, distractors were chosen with embedding similarity $\approx 80\% \times \tau_F$ (see \cref{fig_questionnaire}), making them visually ambiguous.  

\textbf{E. Evaluation Metrics.}

We evaluate privacy leakage using Type I and Type II accuracies~\cite{mai2018reconstruction}, where higher values indicate stronger identity recovery. Type II serves as the primary metric, as it excludes the optimized target image and better reflects practical cross-image identity leakage.

Type I accuracy measures whether a reconstructed image $\hat{x}$ matches the optimized target image $x^{\mathrm{tgt}}$ in the embedding space of a face recognition model $F(\cdot)$:
\begin{equation}
\text{Type I} = \frac{1}{I} \sum_{i=1}^{I} \mathbb{I}\!\left( S(F(\hat{x}_i), F(x^{\mathrm{tgt}}_i)) \geq \tau_F \right),
\end{equation}
where $S(\cdot,\cdot)$ denotes cosine similarity and $\tau_F$ is the threshold.

Type II accuracy evaluates whether $\hat{x}$ matches other images of the same identity:
\begin{equation}
\text{Type II} = \frac{1}{I \times J} \sum_{i=1}^{I} \sum_{j=1}^{J} 
\mathbb{I}\!\left( S(F(\hat{x}_i), F(x^{j \neq \mathrm{tgt}}_i)) \geq \tau_F \right),
\end{equation}
where $J$ is the number of alternative images per identity.

To objectively assess visual quality, we additionally report FID~\cite{heusel2017gans}, LPIPS~\cite{zhang2018unreasonable}, SSIM~\cite{wang2004image}, and NIQE~\cite{mittal2012making}. Lower FID indicates closer alignment with the real-image distribution. Lower LPIPS reflects higher perceptual similarity to ground truth. Higher SSIM indicates better structural fidelity. Lower NIQE suggests fewer unnatural distortions and more natural image statistics. These complementary metrics provide a rigorous evaluation of both identity leakage and visual realism.

\begin{table*}[!t]
    \centering
    \footnotesize
    \begin{threeparttable}
        \caption{Privacy vulnerabilities of face recognition models evaluated under DiffMI.}
        \label{tab_crossmodel}
        \setlength{\tabcolsep}{2.8mm}{\begin{tabular}{cc|cc|cc|cc|cc|cc}
            \toprule
            \multirow{4}{*}{Dataset}&\multirow{4}{*}{\begin{tabular}{c}Target\\Model\end{tabular}}&\multicolumn{10}{c}{Evaluation}\\
            \cmidrule{3-12}
            &&\multicolumn{2}{c|}{FaceNet}&\multicolumn{2}{c|}{ArcFace}&\multicolumn{2}{c|}{DCTDP}&\multicolumn{2}{c|}{PartialFace}&\multicolumn{2}{c}{Average}\\
            \cmidrule{3-12}
            &&Type I&Type II&Type I&Type II&Type I&Type II&Type I&Type II&Type I&Type II\\
            \midrule
            \multirow{4.5}{*}{CelebA}&\textcolor{Green}{FaceNet}&\colorbox{gray!20}{100.00}&\colorbox{gray!20}{96.93}&\ \ 96.70&80.12&\ \ 97.90&84.19&\ \ 94.90&77.67&\textcolor{Green}{97.38}&\textcolor{Green}{84.73}\\
            &\textcolor{red}{ArcFace}&\ \ 99.90&92.76&\colorbox{gray!20}{100.00}&\colorbox{gray!20}{99.06}&100.00&94.23&\ \ 99.90&92.84&\textcolor{red}{99.95}&\textcolor{red}{94.72}\\
            &DCTDP&\ \ 99.30&87.43&\ \ 99.50&89.62&\colorbox{gray!20}{100.00}&\colorbox{gray!20}{96.52}&\ \ 99.90&86.61&99.68&90.05\\
            &PartialFace&\ \ 97.70&80.06&\ \ 99.20&87.29&\ \ 99.60&85.82&\colorbox{gray!20}{100.00}&\colorbox{gray!20}{96.14}&99.13&87.33\\
            \midrule
            \multirow{4.5}{*}{LFW}&\textcolor{Green}{FaceNet}&\colorbox{gray!20}{100.00}&\colorbox{gray!20}{98.56}&\ \ 87.00&66.23&\ \ 93.40&73.31&\ \ 80.80&59.92&\textcolor{Green}{90.30}&\textcolor{Green}{74.51}\\
            &\textcolor{red}{ArcFace}&\ \ 96.00&91.03&\colorbox{gray!20}{100.00}&\colorbox{gray!20}{99.64}&\ \ 99.60&95.88&\ \ 98.60&92.33&\textcolor{red}{98.55}&\textcolor{red}{94.72}\\
            &DCTDP&\ \ 96.70&90.34&\ \ 99.70&95.37&\colorbox{gray!20}{100.00}&\colorbox{gray!20}{99.44}&\ \ 97.80&86.31&\textcolor{red}{98.55}&92.87\\
            &PartialFace&\ \ 90.90&75.39&\ \ 98.70&85.44&\ \ 96.20&77.99&\colorbox{gray!20}{100.00}&\colorbox{gray!20}{98.84}&96.45&84.42\\
            \bottomrule
        \end{tabular}}
        \begin{tablenotes}
            \item \colorbox{gray!20}{Gray cells} indicate evaluation on target models. \textcolor{Green}{Green} and \textcolor{Red}{red} highlight the most and least secure models, with the lowest and highest accuracies, respectively.
            \textbf{Takeaway:} FaceNet~\cite{schroff2015facenet} exhibits the strongest privacy protection, outperforming even inversion-resilient models such as PartialFace~\cite{mi2023privacy} and DCTDP~\cite{ji2022privacy}, while ArcFace~\cite{deng2019arcface} shows the weakest resistance to inversion attacks.
        \end{tablenotes}
    \end{threeparttable}
\end{table*}

\textbf{F. Evaluation Protocol.}
Since model inversion attacks optimize directly against a specific recognition model, outputs may overfit, appearing successful due to adversarial artifacts. We treat such overfitting as failure. To robustly assess privacy risk and reduce model-specific bias, we report \emph{average cross-model accuracy over both target and non-target models}.

As described earlier, we evaluate four face recognition models. In each attack instance, one model serves as the \emph{target} (\ie used for optimization), while the remaining three act as \emph{non-target} models to test generalization. This setup assesses whether identity leakage transfers across architectures.

To ensure consistency and avoid bias from model-specific variability (\eg comparing the average of Models A+B+C with target D \vs B+C+D with target A), we report results \emph{averaged over all four models} (\ie always the average of Models A+B+C+D), regardless of which model is targeted. This provides a more reliable measure of each model’s vulnerability, treating the target model as a biometric system that can also be compromised, rather than excluding it, thereby better reflecting the attack’s generalizability.



\section{Evaluation}
Our method, DiffMI, is evaluated in both white-box (\cref{performance_white}) and black-box (\cref{performance_black}) settings. The Ranked Adversary uses APGD~\cite{croce2020reliable} for white-box optimization and GreedyPixel~\cite{wang2025greedypixel} for black-box fine-grained refinement.

\subsection{White-Box Model Inversion}
\label{performance_white}
\subsubsection{Privacy Vulnerabilities of Face Recognition Models}
We begin by evaluating the privacy vulnerabilities of face recognition models under the proposed DiffMI attack in the white-box setting (setup refers to \cref{tab_settings}). As shown in \cref{tab_crossmodel}, DiffMI successfully compromises all four models, including inversion-resilient variants, with Type~I accuracies ranging from 90.30\% to 99.95\% and Type~II accuracies from 74.51\% to 94.72\% across two datasets.

Interestingly, models explicitly designed for inversion resistance (PartialFace and DCTDP) do not demonstrate superior robustness. In contrast, FaceNet, the oldest and standard model, exhibits the strongest resistance, challenging the effectiveness of current inversion-resilient defenses.

\begin{table}[!t]
    \centering
    \footnotesize
    \begin{threeparttable}
        \caption{User study results with PartialFace~\cite{mi2023privacy} as the target (inversion-resilient) model.}
        \label{tab_userstudy}
        \setlength{\tabcolsep}{5.5mm}{\begin{tabular}{ccc}
            \toprule
            Question&Dataset&Accuracy (\%) $\uparrow$\\
            \midrule
            Same person?&LFW&71.5 (14.3 / 20.0)\\
            Which one matches?&LFW&80.5 (16.1 / 20.0)\\
            Which one matches?&CelebA&81.0 (16.3 / 20.0)\\
            \bottomrule
        \end{tabular}}
    \end{threeparttable}
\end{table}

\textbf{User Study}
To validate that our high quantitative results are not solely due to adversarial overfitting, we conducted a user study to assess whether human perceptual judgments align with the reported Type~II accuracies. Participants were asked two questions: \one “Do you think the reconstruction and target depict the same person?”; and \two “Given three candidates (each with two reference photos), which one matches the reconstruction?” Results in \cref{tab_userstudy} confirm the attack’s practical effectiveness: over 71.5\% of reconstructions were correctly matched to their identities by human participants, further supporting the real-world privacy threat posed by DiffMI.

\begin{table}[!t]
    \centering
    \footnotesize
    \begin{threeparttable}
        \caption{Attack performance of the proposed DiffMI under demographic bias.}
        \label{tab_bias}
        \setlength{\tabcolsep}{3.6mm}{\begin{tabular}{lcc}
            \toprule
            Data Subset&Type I&Type II\\
            \midrule
            Full dataset (100 identities $\times$ 10 images)&98.55&94.72\\
            Racial minority (20 identities $\times$ 10 images)&97.25&93.62\\
            Senior age (20 identities $\times$ 10 images)&99.13&96.62\\
            \bottomrule
        \end{tabular}}
        \begin{tablenotes}
            \item Demographic bias has limited impact, with a maximum drop of $1.3\%$. Results are averaged over four evaluation models using ArcFace~\cite{deng2019arcface} on LFW~\cite{hua2008labeled}.
        \end{tablenotes}
    \end{threeparttable}
\end{table}

\begin{table*}[!t]
    \centering
    \footnotesize
    \begin{threeparttable}
        \caption{Performance of DiffMI and the training-free baseline MAP$^2$V~\cite{zhang2024validating} on four face recognition models.}
        \label{tab_comparison}
        \setlength{\tabcolsep}{1.5mm}{\begin{tabular}{cc|ccc|ccc|ccc|ccc}
            \toprule
            \multirow{2.5}{*}{Dataset}&\multirow{2.5}{*}{Attack}&\multicolumn{3}{c|}{FaceNet}&\multicolumn{3}{c|}{ArcFace}&\multicolumn{3}{c|}{DCTDP}&\multicolumn{3}{c}{PartialFace}\\
            \cmidrule{3-14}
            &&Type I&Type II&Similarity$^3$&Type I&Type II&Similarity$^3$&Type I&Type II&Similarity$^3$&Type I&Type II&Similarity$^3$\\
            \midrule
            \multirow{4.5}{*}{CelebA}&Original$^1$&100.00&97.00&$\tau_C=0.99$&100.00&99.09&$\tau_C=0.99$&100.00&96.62&$\tau_C=0.98$&100.00&95.65&$\tau_C=0.98$\\
            \cmidrule{2-14}
            &Random$^2$&\quad 4.40&\ \ 4.50&0.0822&\quad 1.30&\ \ 1.73&0.0304&\quad 4.20&\ \ 4.74&0.0771&\ \ 10.10&\ \ 9.72&0.1404\\
            &MAP$^2$V&\ \ 89.75&69.19&0.9248&\ \ 95.03&84.61&0.8175&\ \ 96.75&81.12&0.8135&\ \ 97.98&80.84&0.7758\\
            &Ours&\textbf{\ \ 97.38}&\textbf{84.73}&\textbf{0.9920}&\textbf{\ \ 99.95}&\textbf{94.72}&\textbf{0.9898}&\textbf{\ \ 99.68}&\textbf{90.05}&\textbf{0.9818}&\textbf{\ \ 99.13}&\textbf{87.33}&\textbf{0.9818}\\
            \midrule
            \multirow{4.5}{*}{LFW}&Original$^1$&100.00&98.60&$\tau_C=0.99$&100.00&99.62&$\tau_C=0.99$&100.00&99.47&$\tau_C=0.98$&100.00&98.89&$\tau_C=0.98$\\
            \cmidrule{2-14}
            &Random$^2$&\quad 0.70&\ \ 0.91&0.0154&\quad 0.00&\ \ 0.06&0.0046&\quad 0.20&\ \ 0.09&0.0129&\quad 0.50&\ \ 0.54&0.0568\\
            &MAP$^2$V&\ \ 90.05&73.14&0.9337&\ \ 88.98&79.22&0.7723&\ \ 92.50&83.05&0.7787&\ \ 90.54&80.41&0.7636\\
            &Ours&\textbf{\ \ 90.30}&\textbf{74.51}&\textbf{0.9917}&\textbf{\ \ 98.55}&\textbf{94.72}&\textbf{0.9834}&\textbf{\ \ 98.55}&\textbf{92.87}&\textbf{0.9774}&\textbf{\ \ 96.45}&\textbf{84.42}&\textbf{0.9794}\\
            \bottomrule
        \end{tabular}}
        \begin{tablenotes}
            \item $^1$ Upper bound: similarity between ground-truth identity pairs (true target faces). $^2$ Lower bound: similarity from randomly generated faces without targeted reconstruction. $^3$ Average cosine similarity (on the target model) between target embeddings and final reconstructions. Values closer to the confidence threshold $\tau_C$ indicate better convergence. \textbf{Bold} indicates the highest identity recovery performance.
        \end{tablenotes}
    \end{threeparttable}
\end{table*}

\begin{figure}[!t]
    \centering
    \includegraphics[width=\linewidth]{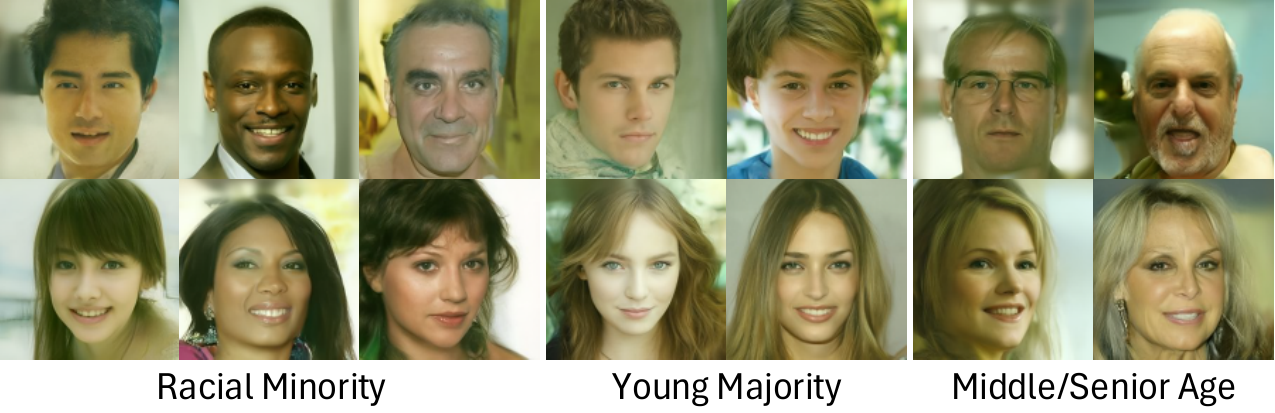}
    \caption{Examples from the generated latent code pool illustrating demographic diversity. Among the 1,000 codes, both racial minority and middle-to-senior age groups are represented. All examples are randomly generated.}
    \label{fig_pool_bias}
\end{figure}

\subsubsection{Concern on Demographic Bias}

DiffMI relies on a diffusion generator pretrained on CelebA-HQ, a dataset known to exhibit demographic imbalance (e.g., race and age distribution). To evaluate whether such imbalance affects attack performance, we conduct additional experiments on two LFW subsets: a racial minority subset and a senior age group subset (each 20 identities$\times$10 images, no overlap). As shown in \cref{tab_bias}, the maximum drop is $1.3\%$ compared to the full dataset, indicating that demographic imbalance in the generator’s training data has limited impact on attack effectiveness.

This limited degradation is largely attributable to the demographic diversity within the 1,000 robust latent codes (see \cref{fig_pool_bias}). Since the diffusion prior models the full training distribution, random sampling naturally preserves its demographic variability. Moreover, Step (b) Top-$N$ Latent Code Selection initializes optimization from the closest candidates in the pool, avoiding the need for drastic demographic transformations (e.g., across age or racial groups). Consequently, demographic imbalance in the generator’s training data has limited impact on attack performance.

In practice, an attacker may retrain or fine-tune the diffusion model to mitigate bias or adapt to a specific domain, which introduces additional computational cost. To quantify this overhead, we retrain the adopted DDPM generator (114M parameters, 455MB) from scratch on CelebA-HQ (256$\times$256 resolution; 28K training and 2K validation images) using a single NVIDIA H100 GPU (96GB memory). The training process requires approximately 16.2 hours. Since the generator can be reused across all target models and subsequent attacks within the same domain, this cost is amortized and remains practically manageable.

\begin{figure}[!t]
    \centering
    \includegraphics[width=\linewidth]{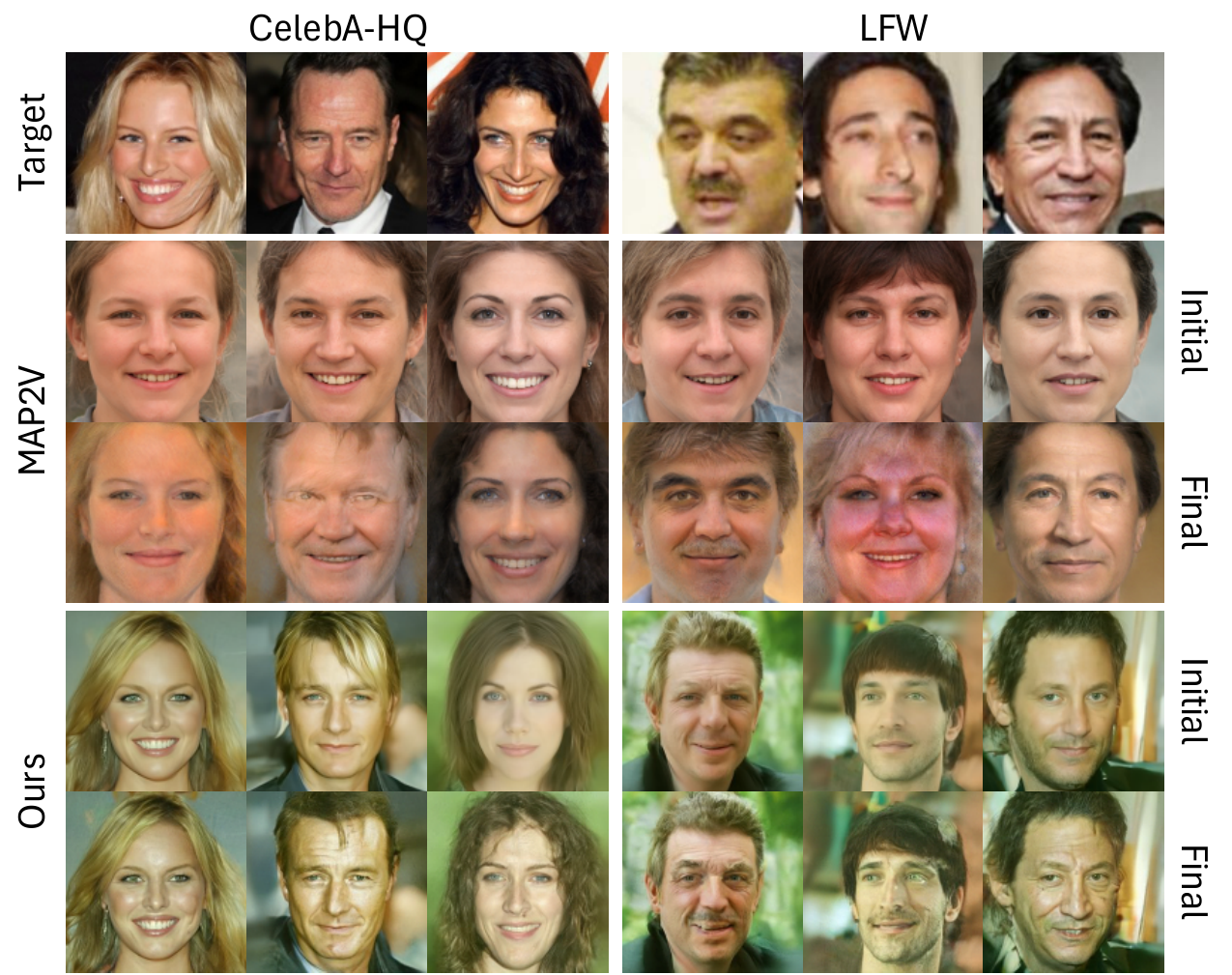}
    \caption{Visual comparison of our DiffMI and MAP$^2$V~\cite{zhang2024validating} in the white-box setting. DiffMI achieves higher identity recovery accuracy and produces full headshot-style reconstructions. ``Initial'' refers to outputs before manipulation. The target model is the inversion-resilient PartialFace~\cite{mi2023privacy}.}
    \label{fig_DiffMI_white}
\end{figure}

\subsubsection{Comparison with Training-Free Baseline}
As summarized in \cref{tab_relatedworks}, existing training-free model inversion methods are all GAN-based. DiffMI represents the first diffusion-driven, training-free approach. We compare DiffMI against the strongest prior method, MAP$^2$V~\cite{zhang2024validating}.

As shown in \cref{tab_comparison}, DiffMI consistently surpasses MAP$^2$V in Type~I and Type~II accuracies and attains higher similarity convergence. This demonstrates that diffusion-based generation enables more accurate identity reconstruction than GANs, despite both being training-free. Moreover, the similarity scores of DiffMI approach the confidence threshold $\tau_C$, a statistically grounded upper bound reflecting the performance of real facial images. This convergence underscores the superior identity fidelity of diffusion models.

We also compare qualitative results in \cref{fig_DiffMI_white}. Visually, DiffMI achieves higher identity resemblance and generates full headshot-style reconstructions, offering finer granularity for identity recovery than MAP$^2$V.

\begin{table*}[!t]
    \footnotesize
    \centering
    \begin{threeparttable}
        \caption{Performance comparison of DiffMI with training-dependent baselines and the naive APGD + DDPM strategy.}
        \label{tab_more_baseline}
        \setlength{\tabcolsep}{1mm}{\begin{tabular}{lcccc|cc|cc|cc|cc|cc}
            \toprule
            \multirow{4}{*}{Attack}&\multirow{4}{*}{Generator}&\multirow{4}{*}{\begin{tabular}{c}Training-\\Free\end{tabular}}&\multirow{4}{*}{Dataset}&\multirow{4}{*}{\begin{tabular}{c}Target\\Model\end{tabular}}&\multicolumn{10}{c}{Evaluation}\\
            \cmidrule{6-15}
            &&&&&\multicolumn{2}{c|}{FaceNet}&\multicolumn{2}{c|}{ArcFace}&\multicolumn{2}{c|}{DCTDP}&\multicolumn{2}{c|}{PartialFace}&\multicolumn{2}{c}{Avg.}\\
            \cmidrule{6-15}
            &&&&&Type I&Type II&Type I&Type II&Type I&Type II&Type I&Type II&Type I&Type II\\
            \midrule
            
            DSCasConv~\cite{shahreza2024vulnerability}&DeconvNet&$\usym{2717}$&\multirow{7}{*}{LFW}&\multirow{7}{*}{ArcFace}&\textbf{97.10}&\underline{90.42}&\colorbox{gray!20}{100.00}&\colorbox{gray!20}{99.03}&\textbf{99.90}&\textbf{97.16}&\textbf{99.50}&\textbf{93.21}&\textbf{99.13}&\textbf{94.96}\\
            Shahreza~\etal \cite{shahreza2024template}&StyleGAN&$\usym{2717}$&&&61.00&49.33&\colorbox{gray!20}{\ \ 98.10}&\colorbox{gray!20}{77.79}&84.10&59.41&74.00&52.24&79.30&59.69\\
            MAP$^2$V \cite{zhang2024validating}&StyleGAN&$\usym{2713}$&&&67.60&60.40&\colorbox{gray!20}{100.00}&\colorbox{gray!20}{99.42}&97.10&83.78&91.20&73.27&88.98&79.22\\
            APGD (Random)$^*$&DDPM&$\usym{2713}$&&&69.10&61.33&\colorbox{gray!20}{100.00}&\colorbox{gray!20}{99.56}&92.30&75.40&84.50&66.51&86.48&75.70\\
            APGD (Top 1/1000)$^*$&DDPM&$\usym{2713}$&&&78.00&71.70&\colorbox{gray!20}{100.00}&\colorbox{gray!20}{\underline{99.63}}&96.00&83.91&90.70&76.59&91.18&82.96\\
            DiffMI (Ours)&DDPM&$\usym{2713}$&&&\underline{96.00}&\textbf{91.03}&\colorbox{gray!20}{100.00}&\colorbox{gray!20}{\textbf{99.64}}&\underline{99.60}&\underline{95.88}&\underline{98.60}&\underline{92.33}&\underline{98.55}&\underline{94.72}\\
            \bottomrule
        \end{tabular}}
        \begin{tablenotes}
            \item \colorbox{gray!20}{Gray cells} indicate evaluation on target models. \textbf{Bold} and \underline{underline} denote the highest and second-highest identity recovery performance, respectively. $^*$~Direct APGD attacks in the DDPM latent space without additional strategies. ``Top 1/1000'' selects the best of 1{,}000 samples; ``Random'' uses one unfiltered sample.
            \textbf{Takeaway:} 
            \one The training-dependent DeconvNet-based method~\cite{shahreza2024vulnerability} achieves slightly higher recovery accuracy than our DiffMI but requires \textbf{6.4 million queries}, \textbf{1.5 days} of training on \textbf{two A100 80GB GPUs}, and produces a target-specific generator that generalizes only to ArcFace. 
            \two Both DiffMI and the training-heavy DeconvNet significantly outperform GAN-based methods and the naive APGD + DDPM approach. 
            \three The naive APGD + DDPM strategy exhibits severe overfitting to the target model, achieving near-perfect performance on the target but a substantial drop on non-target evaluations.
        \end{tablenotes}
    \end{threeparttable}
\end{table*}

\subsubsection{Comparison with Training-Dependent Baselines}
Although our proposed DiffMI is training-free, making it significantly more computationally efficient and broadly applicable than training-dependent approaches, we include comparisons for completeness. As summarized in \cref{tab_relatedworks}, most training-dependent model inversion attacks rely on DeconvNets or GANs. We select two recent representatives: a DeconvNet-based method~\cite{shahreza2024vulnerability} and a GAN-based method~\cite{shahreza2024template}.

As shown in \cref{tab_more_baseline}, despite being training-free, DiffMI matches the recovery performance of the DeconvNet-based method~\cite{shahreza2024vulnerability} and significantly outperforms the GAN-based method~\cite{shahreza2024template}. The DeconvNet approach attains a slightly higher Type~II accuracy (0.24\% above DiffMI) but at substantial cost: training its target-specific generator requires \textbf{1.5 days} on \textbf{two A100 80GB GPUs} and \textbf{6.4 million queries} for gradient computation (ours costs 98--256 seconds and 1,300 queries per image for any model). Moreover, the generator is usable only for the target model it was trained on (ArcFace in \cref{tab_more_baseline}); for any new target, the entire costly training process must be repeated. In contrast, DiffMI is fully training-free and applies directly to any face recognition model, requiring only 98--256 seconds and 1,300 queries per image.

\begin{figure}[!t]
    \centering
    \includegraphics[width=\linewidth]{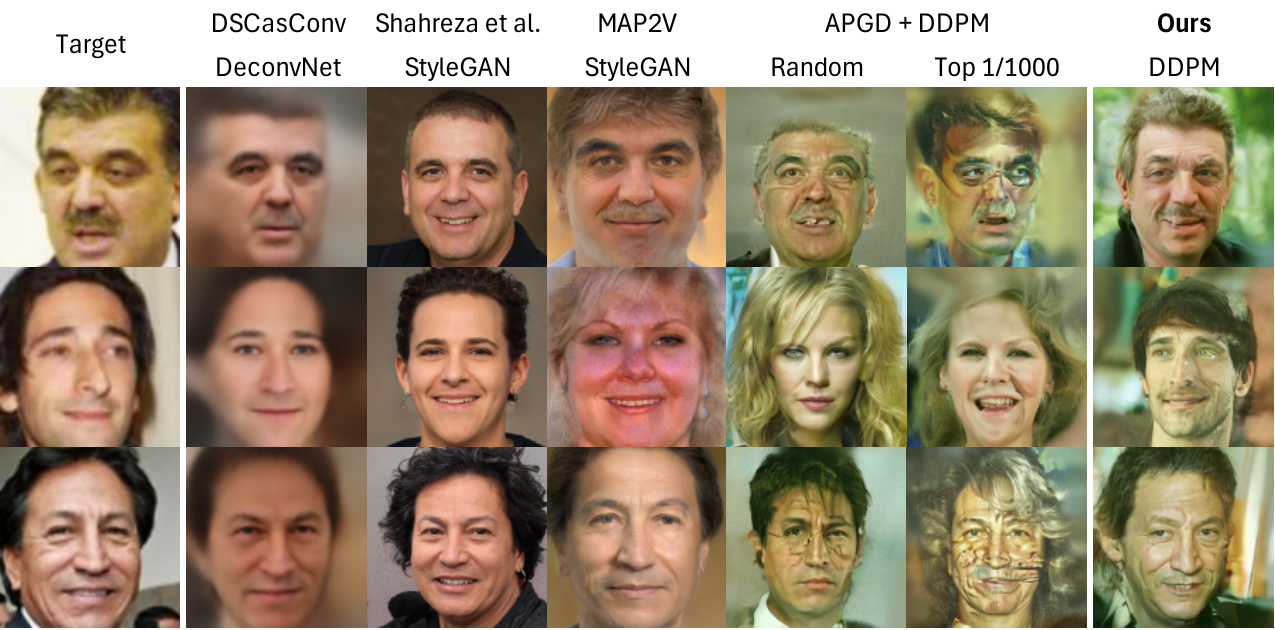}
    \caption{Visual fidelity comparison across all baselines. Both Shahreza~\etal~\cite{shahreza2024template} and our method produce high-resolution, headshot-style reconstructions with strong perceptual quality. In contrast, DSCasConv~\cite{shahreza2024vulnerability} yields low-resolution, blurry outputs with diminished facial detail. Naively applying APGD in the DDPM latent space results in severe artifacts, even when initialized from the top-1 latent code among 1,000 candidates.}
    \label{fig_compare}
\end{figure}

Despite its computational burden, the DeconvNet method is limited to generating low-resolution ($112 \times 112$) images. Furthermore, as illustrated in \cref{fig_compare}, it produces the lowest visual fidelity among all methods, yielding blurry and low-detail reconstructions. These limitations underscore the practical advantages of training-free methods, especially those leveraging expressive and resolution-flexible generative backbones such as GANs or diffusion models.

\subsubsection{Comparison with Naive APGD + DDPM}
A straightforward approach to applying diffusion models for training-free model inversion is to directly run adversarial attacks (\eg APGD~\cite{croce2020reliable}) in the latent space of an unconditional DDPM. However, this naive combination leads to severe artifacts and overfitting on the target model. These concerns are empirically validated in \cref{tab_more_baseline,fig_compare}.

As shown in \cref{tab_more_baseline}, the naive ``APGD + DDPM" strategy attains near-perfect Type~II accuracy on the target model (approaching 100\%) but suffers large drops on non-target models, ranging from $-27.93\%$ to $-15.72\%$. In contrast, our controlled diffusion manipulation method limits this gap to $-8.61\%$ to $-3.76\%$, demonstrating superior robustness and transferability. Notably, non-target model results are typically lower than those of the target model.

Moreover, applying APGD directly in the DDPM latent space introduces severe artifacts (\cref{fig_compare}), even when initialized from the top-1 latent code among 1,000 candidates. These results further support the necessity of principled latent manipulation rather than naive adversarial optimization.

\begin{figure*}[!t]
    \centering
    \includegraphics[width=\linewidth]{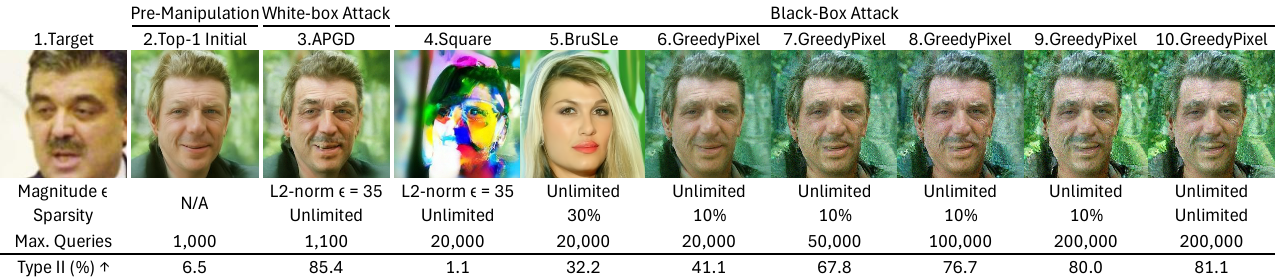}
    \caption{Performance of DiffMI in the black-box setting. Compared to its white-box counterpart (columns 3 \vs 10), black-box DiffMI achieves slightly lower Type~II accuracy. Among black-box methods (columns 4–6), only the fine-grained GreedyPixel algorithm successfully manipulates the diffusion model’s latent codes. Columns 6–10 show that increasing the attack budget (queries and sparsity) improves inversion performance. Notably, 40\% of identities become recognizable after just 20,000 queries, with diminishing returns beyond 100,000. The dataset is LFW, the target model is the inversion-resilient PartialFace~\cite{mi2023privacy}, and evaluation is performed using ArcFace~\cite{deng2019arcface}.}
    \label{fig_DiffMI_black}
\end{figure*}

\subsection{Black-Box Model Inversion}
\label{performance_black}
We evaluate DiffMI in the black-box setting under various adversarial attack algorithms and compare its performance to the white-box variant. As shown in \cref{fig_DiffMI_black} (columns 3 \vs 10), GreedyPixel achieves Type~II accuracy slightly lower than white-box APGD at higher query cost.

\cref{fig_DiffMI_black} (columns 4–6) further demonstrates that only fine-grained black-box methods such as GreedyPixel can effectively manipulate the diffusion model's latent codes. In contrast, coarser strategies like Square~\cite{andriushchenko2020square} and BruSLe~\cite{vo2024brusleattack} fail to produce successful reconstructions.

Finally, comparing columns 6–10, increasing the query budget and relaxing sparsity constraints improves inversion performance by allowing modification of more latent dimensions, yielding more accurate identity recovery.

\subsection{Analysis of Identity Overlap Between Private Data and Generator Training Data}

In practice, diffusion generators are trained on large-scale public datasets that are expected to contain none or only a small fraction of the private identities being attacked. In our setting, the generator is pretrained on CelebA-HQ, while LFW serves as the primary evaluation dataset. To quantify potential identity overlap, we conduct an explicit analysis. \textbf{Private data:} LFW (100 identities $\times$ 10 images) with publicly available names. \textbf{Generator training data:} CelebA-HQ (30K images) without identity annotations. Using embedding-based matching followed by manual verification, we find that 8 out of 100 LFW identities appear in CelebA-HQ. In total, 26 images (approximately 0.09\% of the 30K training images) correspond to these identities, indicating minimal overlap.

We further evaluate whether such overlap improves inversion performance by selecting 1,000 CelebA-HQ images as attack targets (i.e., within the generator’s training domain). As shown in \cref{tab_crossmodel}, distribution alignment yields only marginal gains and occasionally slight degradation (e.g., +2.82\% accuracy for DCTDP on LFW versus CelebA-HQ). This suggests that, in the training-free open-set setting, identity overlap provides limited benefit and does not drive attack effectiveness.

\begin{figure}[!t]
    \centering
    \includegraphics[width=\linewidth]{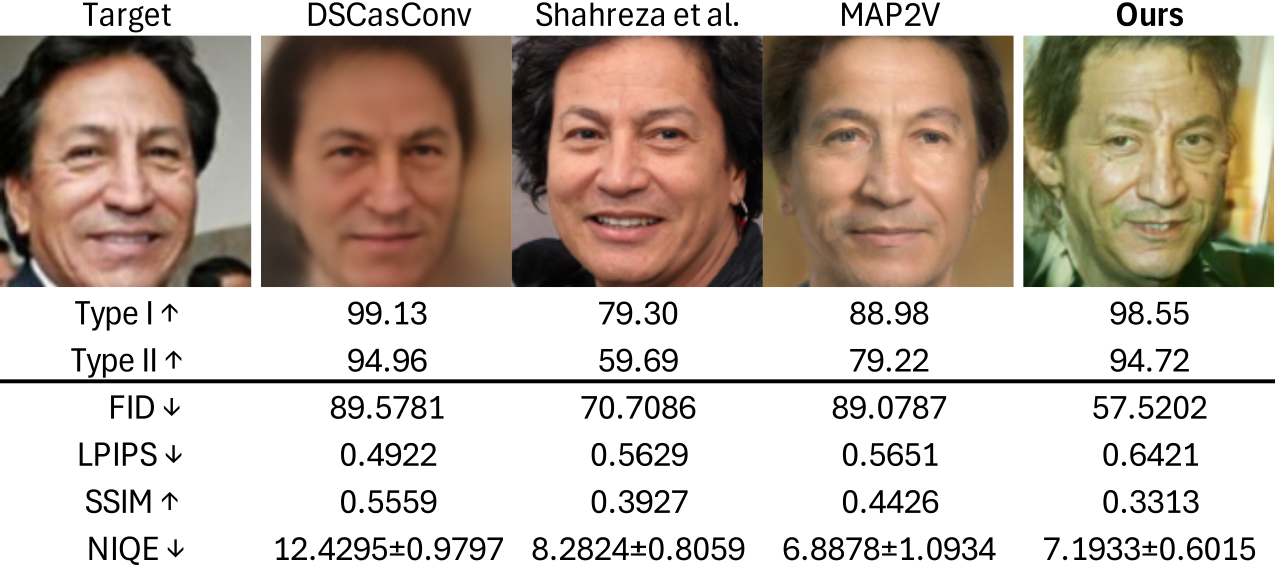}
    \caption{Visual quality comparison on LFW~\cite{hua2008labeled} with ArcFace~\cite{deng2019arcface} as the target model. Despite high identity matching accuracy, FID, LPIPS, and SSIM indicate that embedding-based inversion does not pursue pixel-level reconstruction.}
    \label{fig_quality_eval}
\end{figure}


\subsection{Qualitative Analysis}

As discussed in \cref{problem_loss}, embedding-based model inversion targets identity recovery rather than exact pixel-level reconstruction. \cref{fig_quality_eval} reports FID, LPIPS, SSIM, and NIQE to provide an assessment of visual quality under identical preprocessing (same MTCNN-based cropping with face size 160 and margin 32). Despite high identity matching accuracy, FID, LPIPS, and SSIM confirm that embedding-based methods do not pursue pixel-level replication. In contrast, NIQE shows that DiffMI introduces the fewest unnatural distortions and best naturalness among compared approaches, indicating improved naturalness while preserving identity consistency.

\emph{Identity consistency vs. pixel-level reconstruction:} Face recognition systems operate in embedding space, where identity is determined by similarity rather than pixel correspondence. Thus, generating any image recognized as the same individual constitutes a privacy breach, even if visual details differ. Moreover, models such as ArcFace and FaceNet compress high-dimensional images into low-dimensional embeddings (e.g., 512D), discarding non-identity information and rendering exact inversion ill-posed. In realistic threat scenarios (e.g., re-identification or impersonation), recovering identity—not reproducing a specific photograph—is the primary objective. DiffMI therefore emphasizes identity-consistent reconstruction, aligning with both theoretical properties of embedding-based systems and practical privacy risks.

\section{Ablation Study}
We conduct ablation studies on DiffMI using the LFW~\cite{hua2008labeled} dataset with inversion-resilient PartialFace~\cite{mi2023privacy} as the target model. Only Type~II accuracy is reported, as it offers a more rigorous evaluation than Type~I.

\begin{table}[!t]
    \footnotesize
    \centering
    \begin{threeparttable}
        \caption{Performance (\%) with and without the confidence-aware objective.}
        \label{tab_ablation_stop}
        \setlength{\tabcolsep}{6.1mm}{\begin{tabular}{lcc}
            \toprule
            Confidence-Aware&Time (s) $\downarrow$&Type II (\%) $\uparrow$\\
            \midrule
            $\usym{2713}$ (ours)&\textbf{256}&84.42\\
            $\usym{2717}$&530&\textbf{85.31}\\
            \bottomrule
        \end{tabular}}
        \begin{tablenotes}
            \item \textbf{Bold} indicates the best performance. The confidence-aware objective reduces runtime by over 50\% with marginal accuracy loss.
        \end{tablenotes}
    \end{threeparttable}
\end{table}

\begin{table}[!t]
    \footnotesize
    \centering
    \begin{threeparttable}
        \caption{Ablation of the confidence threshold $\tau_C$.}
        \label{tab_ablation_margin}
        \setlength{\tabcolsep}{2.9mm}{\begin{tabular}{lccc}
            \toprule
            Threshold $\tau_C$&Time (s) $\downarrow$&Similarity $\uparrow$&Type II (\%) $\uparrow$\\
            \midrule
            0.70&\textbf{\ 23}&0.7294&69.93\\
            0.80&\ 38&0.8204&77.59\\
            0.90&\ 64&0.9111&83.11\\
            0.97&213&0.9712&84.23\\
            0.98 (from \cref{eq_tauC})&256&0.9794&84.42\\
            0.99&442&\textbf{0.9861}&\textbf{84.98}\\
            \bottomrule
        \end{tabular}}
        \begin{tablenotes}
            \item \textbf{Bold} indicates the best observed value per metric.
        \end{tablenotes}
    \end{threeparttable}
\end{table}

\subsection{Confidence-Aware Objective}
\label{ablation_margin}
We incorporate a confidence-aware objective into the ranked adversary strategy, enabling early stopping once a reconstruction exceeds a defined confidence threshold. This prioritizes highly-aligned latent codes and avoids unnecessary computation. As shown in \cref{tab_ablation_stop}, the proposed objective significantly reduces runtime with only a minor decrease in Type~II accuracy, offering an effective trade-off between efficiency and attack performance.

The confidence threshold $\tau_C$, defined in \cref{eq_tauC}, specifies the similarity required for early termination. It serves as a soft upper bound, statistically estimated from real facial images. To ensure robustness, $\tau_C$ should be significantly greater than the decision threshold $\tau_F$, but not excessively high (\eg $\tau_C = 1$), which may increase computational cost.

As shown in \cref{tab_ablation_margin}, lower $\tau_C$ values (\eg $<0.90$) yield faster attacks but reduce identity recovery, whereas higher values (\eg $\tau_C > 0.98$) offer marginal accuracy gains at substantially higher runtime. Empirically, $\tau_C = 0.98$ (from \cref{eq_tauC}) provides a strong balance between performance and efficiency. DiffMI is not highly sensitive to $\tau_C$; even $\tau_C = 0.90$ offers a reasonable trade-off. The threshold should avoid being set too low or too close to $1.0$, with \cref{eq_tauC} serving as a statistical guideline rather than a strict requirement.

\begin{figure}[!t]
    \centering
    \includegraphics[width=\linewidth]{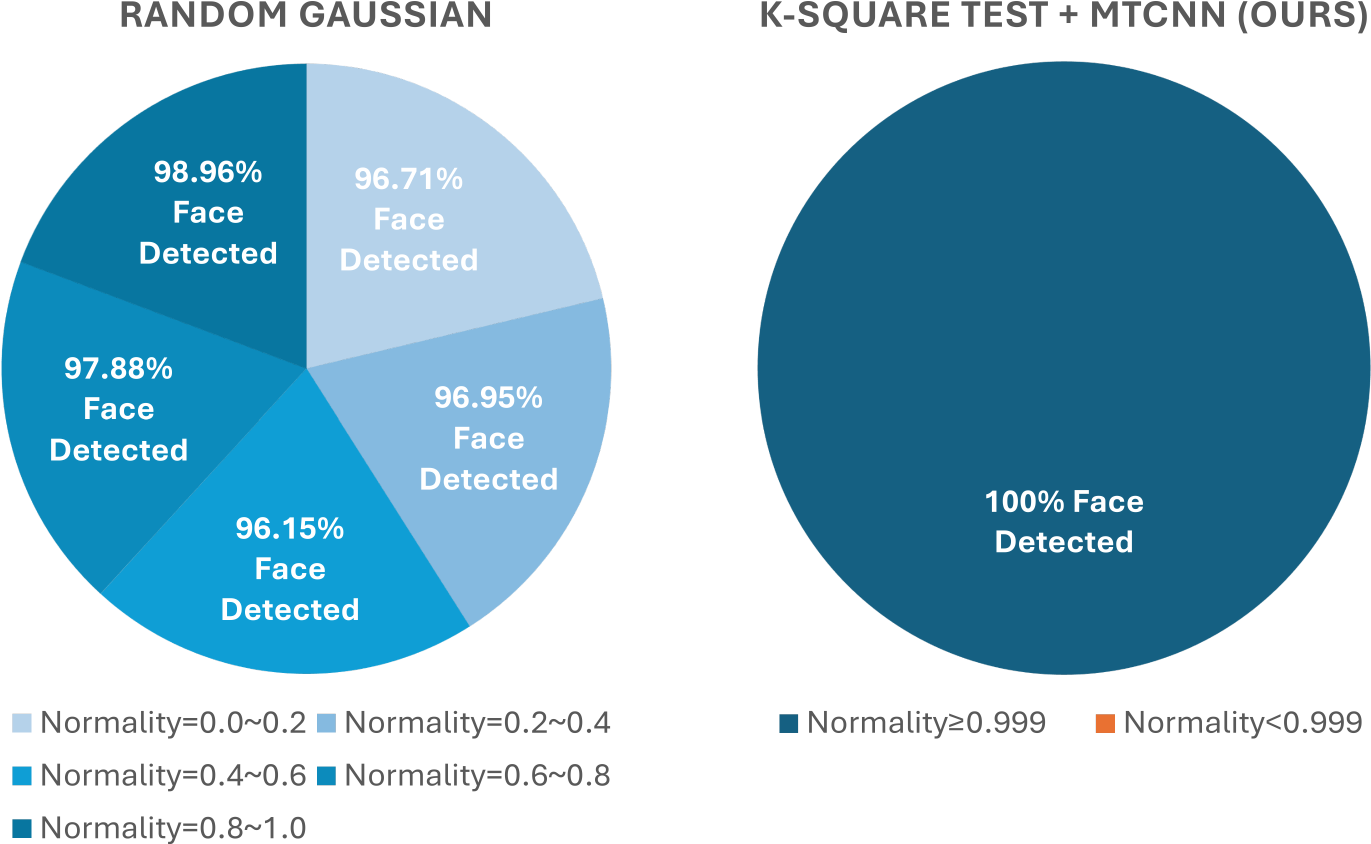}
    \caption{Quality of randomly generated latent codes and effectiveness of the proposed selection strategy. In the left subfigure, Gaussian normality varies across samples, with higher normality generally correlating with improved face detection rates (darker regions). In the right subfigure, our two-stage strategy enforces 100\% $p_K \geq 0.999$ and consistently achieves a 100\% face detection rate ($p_D \geq 0.999$).}
    \label{fig_normality}
\end{figure}

\begin{figure}[!t]
    \centering
    \includegraphics[width=3.2in]{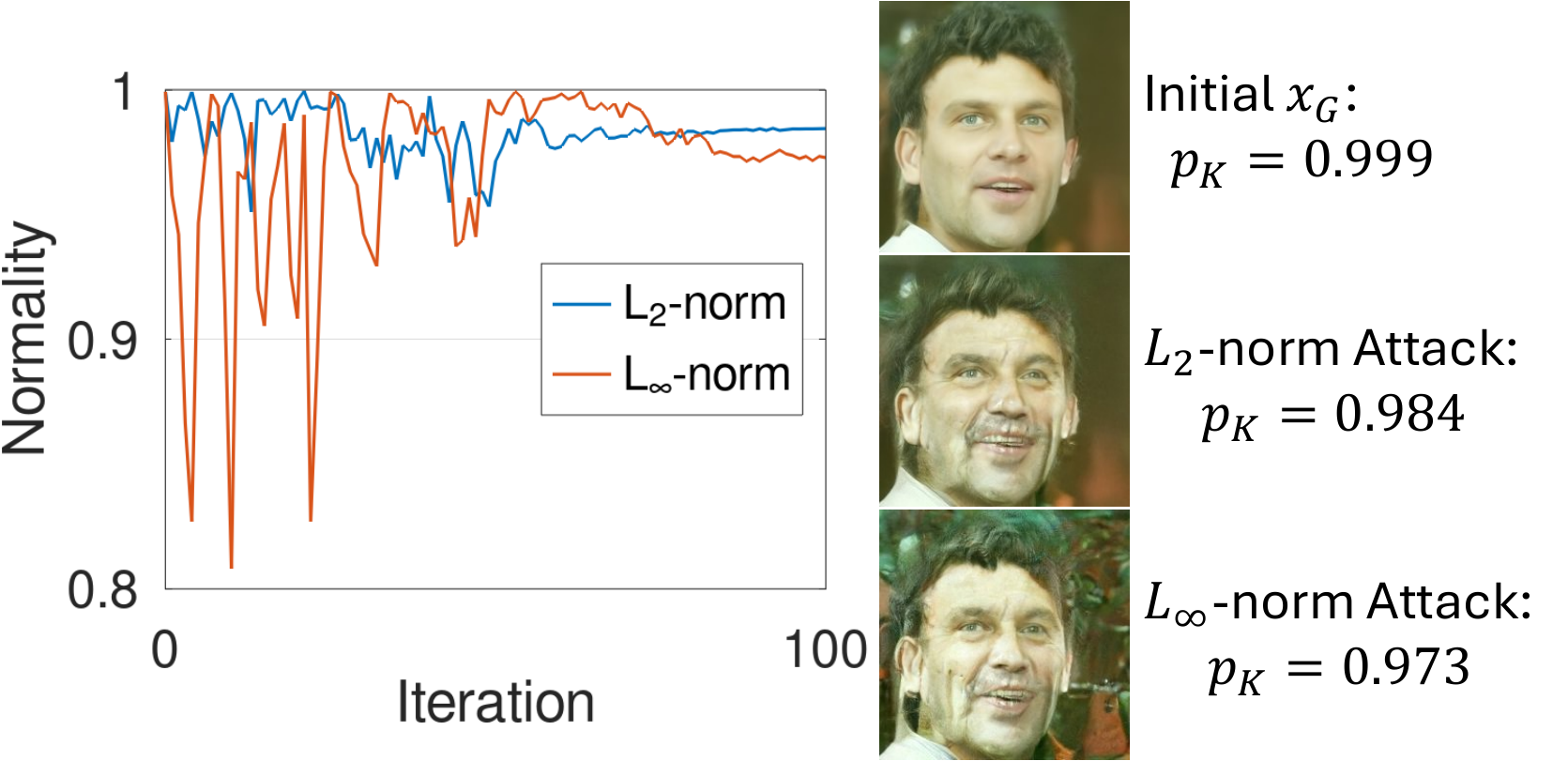}
    \caption{Impact of adversarial manipulation on Gaussian normality. $L_2$-constrained adversaries better preserve latent normality than $L_\infty$-constrained ones, resulting in improved reconstruction fidelity.}
    \label{fig_normality_gap}
\end{figure}

\begin{figure}[!t]
    \centering
    \includegraphics[width=\linewidth]{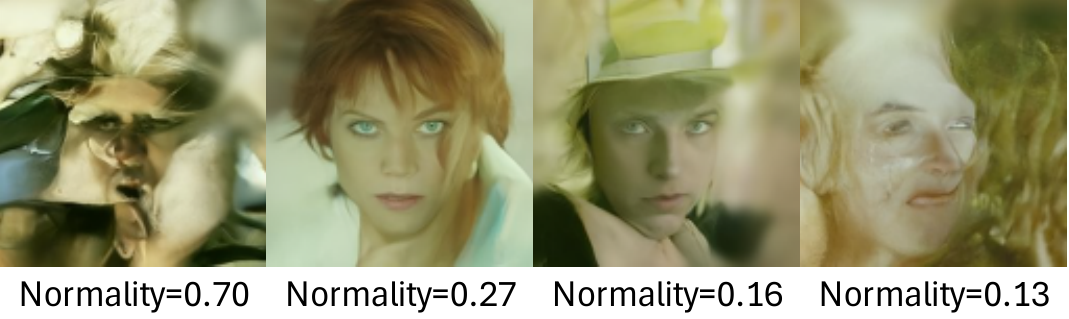}
    \caption{Examples of failed face generation from randomly sampled latent codes. MTCNN fails to detect a face, resulting in detection confidence $p_D = 0$.}
    \label{fig_highnormality_fail}
\end{figure}

\begin{table}[!t]
    \footnotesize
    \centering
    \begin{threeparttable}
        \caption{Ablation of latent code selection strategies.}
        \label{tab_ablation_latentcode}
        \setlength{\tabcolsep}{4.8mm}{\begin{tabular}{ccc}
            \toprule
            Strategy&Time (s) $\downarrow$&Type II (\%) $\uparrow$\\
            \midrule
            Random Gaussian&290&82.78\\
            $K^2$ Test&293&82.47\\
            $K^2$ Test + MTCNN (ours)&\textbf{256}&\textbf{84.42}\\
            \bottomrule
        \end{tabular}}
        \begin{tablenotes}
            \item \textbf{Bold} indicates the superiority of our strategy.
        \end{tablenotes}
    \end{threeparttable}
\end{table}

\subsection{Robustness of Latent Codes}  
\label{ablation_latent}
As illustrated in \cref{fig_normality}, the normality of randomly sampled codes varies, with higher normality generally correlating with higher face detection rates. However, as shown in \cref{fig_normality_gap}, even highly normal codes can lose normality after adversarial manipulation, with greater distortion leading to more severe visual artifacts. Moreover, as shown in \cref{fig_highnormality_fail}, latent codes satisfying Gaussian normality may still fail to produce recognizable faces. These distortion and failure cases underscore the importance of our strategy in guaranteeing optimal initialization, ensuring that 100\% of samples meet $p_K \geq 0.999$ and all facial regions are detected. As shown in \cref{tab_ablation_latentcode}, our strategy, combining $K^2$ test with MTCNN face detection, produces the most robust latent codes, improving Type~II accuracy and reducing runtime.

Furthermore, \cref{tab_tauKD} reports the acceptance rate during robust pool construction. Stricter thresholds reduce the retention ratio but increase computation; under $\tau_K = \tau_D = 0.999$, only 0.06\% of samples are retained. However, this pool is constructed once (approximately 2.7 hours) and reused across all target models and attacks, making the cost fully amortized and practically manageable.

\begin{table*}[!t]
    \centering
    \footnotesize
    \begin{threeparttable}
        \caption{Effect of the latent filtering threshold $\tau_K$ and $\tau_D$ on candidate retention ratio and computational cost.}
        \label{tab_tauKD}
        \setlength{\tabcolsep}{1.6mm}
        \begin{tabular}{lcccccccccc}
            \toprule
            \multirow{2}{*}{Step} 
            & \multicolumn{2}{c}{$\tau_K = \tau_D = 0$ (Random)} 
            & \multicolumn{2}{c}{$\tau_K = \tau_D = 0.8$} 
            & \multicolumn{2}{c}{$\tau_K = \tau_D = 0.9$} 
            & \multicolumn{2}{c}{$\tau_K = \tau_D = 0.99$} 
            & \multicolumn{2}{c}{$\tau_K = \tau_D = 0.999$ (Ours)} \\
            \cmidrule(lr){2-3} \cmidrule(lr){4-5} \cmidrule(lr){6-7} \cmidrule(lr){8-9} \cmidrule(lr){10-11}
            & Ratio (\%) & Time (s)
            & Ratio (\%) & Time (s)
            & Ratio (\%) & Time (s)
            & Ratio (\%) & Time (s)
            & Ratio (\%) & Time (s) \\
            \midrule
            Step 1: Normality Test 
                & 100.00 & 8 
                & 17.55 & 26 
                & 8.57 & 43 
                & 0.87 & 381 
                & 0.08 & 6127 \\
            Step 2: Face Detection 
                & 100.00 & 990 
                & 99.80 & 1008 
                & 99.80 & 1001 
                & 98.81 & 1187 
                & 74.79 & 3657 \\
            \midrule
            Total 
                & 100.00 & 998 
                & 17.51 & 1034 
                & 8.56 & 1044 
                & 0.86 & 1569 
                & 0.06 & 9784 \\
            \bottomrule
        \end{tabular}
    \end{threeparttable}
    \begin{tablenotes}
        \item Stricter thresholds reduce retention but increase computation; under the strictest setting, the one-time pool construction takes approximate 2.7 hours and is reused for all subsequent attacks, making the cost amortized and practical.
        \end{tablenotes}
\end{table*}

\subsection{Top-N Selection}
\begin{figure}[!t]
    \centering
    \includegraphics[width=\linewidth]{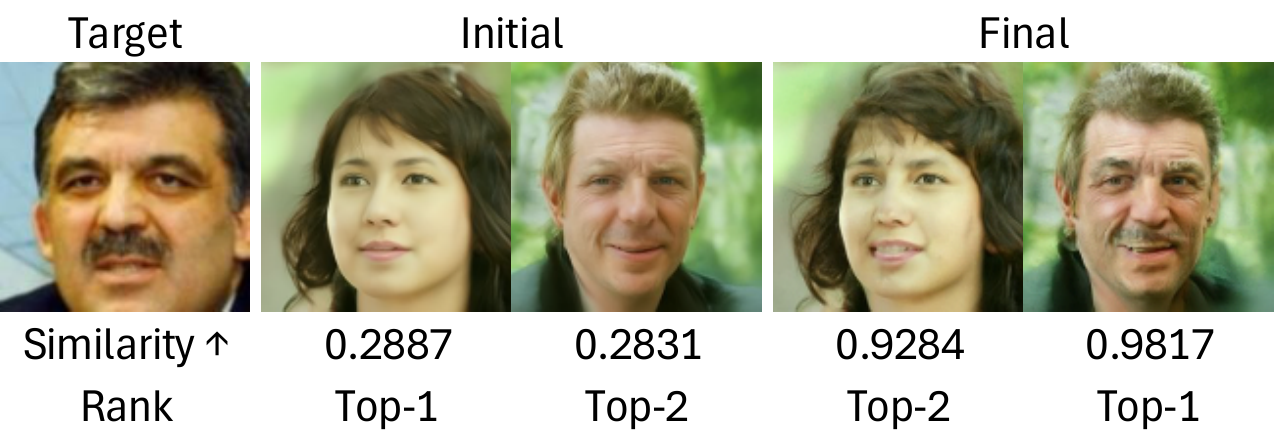}
    \caption{Example where the latent code with the best initial reconstruction fails to achieve the highest final similarity after adversarial optimization.}
    \label{fig_TopN_fail}
\end{figure}

As demonstrated in \cref{fig_TopN_fail}, selecting the code with highest initial similarity can underperform after adversarial optimization. Therefore, DiffMI selects the top $N$ latent codes, improving efficiency through a ranked adversary strategy without processing all candidates. As shown in \cref{tab_ablation_TopN}, larger $N$ increases Type~II accuracy at the cost of longer runtime.

\begin{table}[!t]
    \footnotesize
    \centering
    \begin{threeparttable}
        \caption{Ablation of the top-$N$ value in latent code selection.}
        \label{tab_ablation_TopN}
        \setlength{\tabcolsep}{8.5mm}{\begin{tabular}{ccc}
            \toprule
            Top $N$&Time (s) $\downarrow$&Type II (\%) $\uparrow$\\
            \midrule
            1&\textbf{189}&81.06\\
            3&256&84.42\\
            5&289&\textbf{85.21}\\
            \bottomrule
        \end{tabular}}
        \begin{tablenotes}
            \item \textbf{Bold} indicates the best performance. Increasing $N$ improves reconstruction quality but incurs higher computational cost.
        \end{tablenotes}
    \end{threeparttable}
\end{table}

\subsection{Perturbation Constraint}
As shown in \cref{fig_normality_gap}, $L_2$-norm-constrained adversaries better preserve distributional normality compared to their $L_\infty$ counterparts, resulting in higher-fidelity reconstructions. This is supported by \cref{tab_ablation_constraint}, where $L_2$ and $L_\infty$ attacks yield similar similarity scores, but the $L_2$ constraint ($\epsilon=35$) achieves substantially higher accuracy.

\begin{figure}[!t]
    \centering
    \includegraphics[width=2.1in]{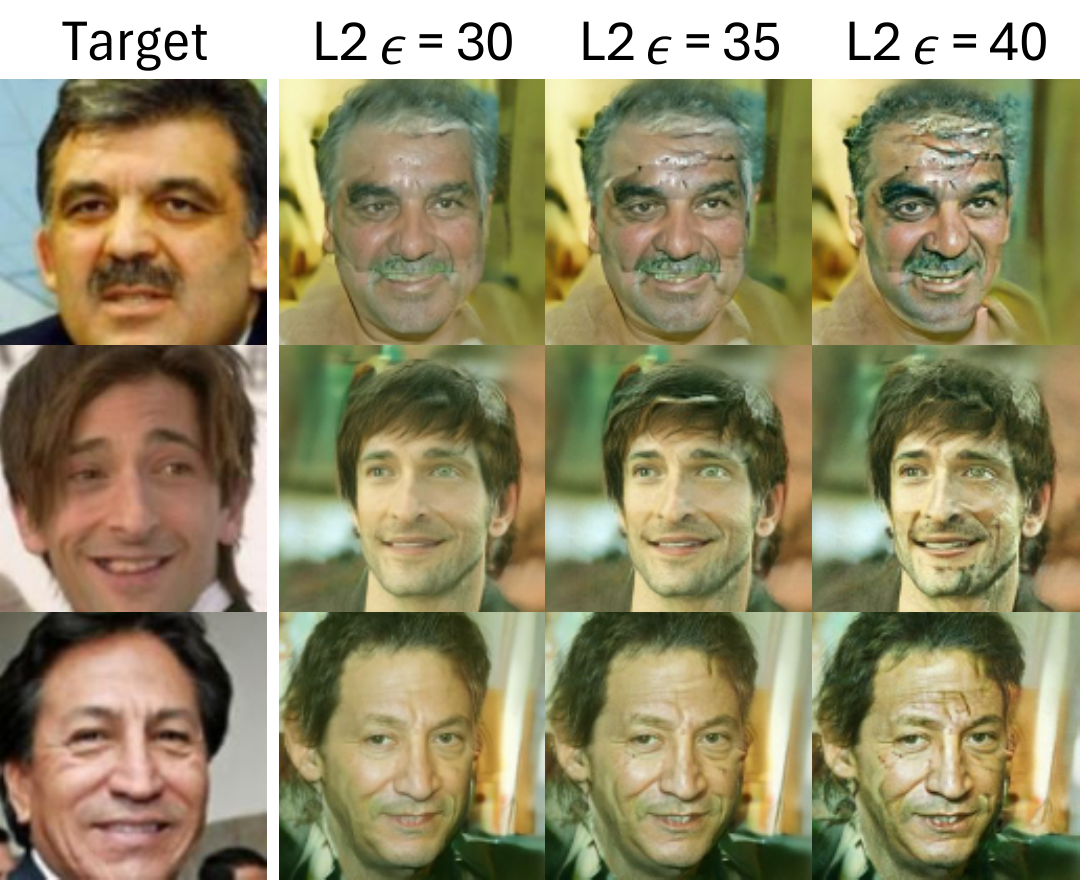}
    \caption{Larger perturbations introduce stronger artifacts, increasing visible noise and reducing generation realism.}
    \label{fig_ablation_eps}
\end{figure}

While \cref{tab_ablation_constraint} suggests that increasing $\epsilon$ leads to improved accuracy and reduced runtime, \cref{fig_ablation_eps} reveals a trade-off: larger perturbation magnitudes introduce stronger visual artifacts, degrading perceptual quality. Thus, the optimal $\epsilon$ must balance reconstruction effectiveness with visual naturalness.

\begin{table}[!t]
    \footnotesize
    \centering
    \begin{threeparttable}
       \caption{Ablation of perturbation constraints across different norms and magnitudes.}
        \label{tab_ablation_constraint}
        \setlength{\tabcolsep}{2.4mm}{\begin{tabular}{ccccc}
            \toprule
            Norm&Magnitude $\epsilon$&Time (s) $\downarrow$&Similarity $\uparrow$&Type II (\%) $\uparrow$\\
            \midrule
            $L_2$&30&455&0.9707&81.49\\
            $L_2$&35&256&0.9794&84.42\\
            $L_2$&40&\textbf{191}&\textbf{0.9816}&\textbf{86.11}\\
            $L_\infty$&0.15&260&0.9791&75.28\\
            \bottomrule
        \end{tabular}}
        \begin{tablenotes}
            \item \textbf{Bold} indicates the best result per metric.
        \end{tablenotes}
    \end{threeparttable}
\end{table}

\section{Conclusion}
We present DiffMI, the first diffusion-driven, training-free model inversion attack against embedding-based face recognition systems. Unlike prior methods relying on GANs or target-specific training, DiffMI manipulates latent codes of an unconditional diffusion model to produce high-fidelity, identity-preserving reconstructions. Extensive white-box and black-box evaluations show that DiffMI compromises diverse face recognition models, including those designed for inversion resistance. It consistently outperforms existing training-free baselines and approaches the performance of training-dependent attacks, without target-specific training or generator tuning. A user study further validates the perceptual realism of our reconstructions.
Our findings suggest that model inversion is inherently difficult to defend due to its reliance on inference-time behavior. If a face recognition system excels at identity matching, it may be intrinsically vulnerable to inversion. Future work should explore stronger defenses against training-free attacks, particularly those leveraging high-fidelity generative priors. The safe deployment of such models in privacy-sensitive applications also warrants deeper study, including robust inversion detection, mitigation strategies, and privacy-preserving architectures.
\section*{Ethical Considerations}

Model inversion attacks on embedding-based face recognition systems have dual implications. On the positive side, strong inversion methods provide a rigorous worst-case evaluation of privacy leakage, enabling systematic assessment of the risks inherent in biometric pipelines and informing the design of more robust defenses. On the negative side, high-fidelity reconstruction from embeddings may enable identity misuse or cross-system privacy leakage if deployed maliciously. Given that facial data constitutes sensitive biometric information under regulations such as GDPR and CCPA, such risks must be carefully considered. All experiments in this work use publicly available datasets with appropriate consent. No proprietary systems are targeted, and no identities outside these datasets are reconstructed. The purpose of this study is to advance privacy auditing and strengthen protective mechanisms in face recognition systems.
{
    \bibliographystyle{ieeetr}
    \bibliography{main}
}
\begin{IEEEbiography}[{\includegraphics[width=1in,height=1.25in,clip,keepaspectratio]{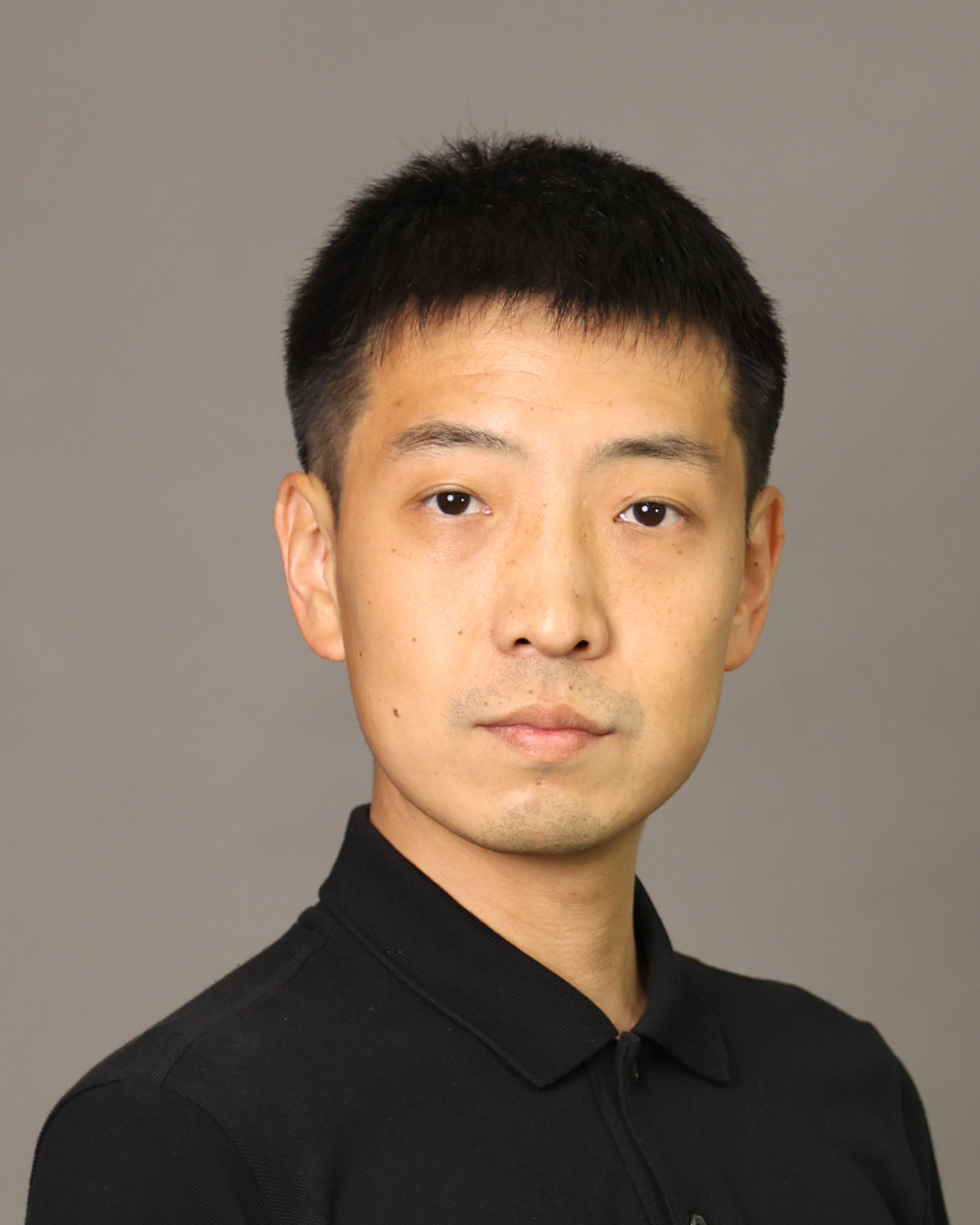}}]{Hanrui Wang}
received his B.S. degree in Electronic Information Engineering from Northeastern University (China) in 2011. After working in the IT industry and rising to a director-level position, he transitioned to academic research in 2019 and earned his Ph.D. in Computer Science from Monash University, Australia, in January 2024. He is currently an Assistant Professor in the Echizen Laboratory at the National Institute of Informatics (NII) in Tokyo, Japan. His research focuses on AI security and privacy, with emphasis on adversarial machine learning, model inversion, and trustworthy multimedia systems. He has published multiple papers in leading journals such as IEEE TIFS, IEEE TDSC, ACM TOMM, and J-STARS, as well as in reputable international conferences including WWW, WACV, FG, ICPR, ICASSP, and ICMI.
\end{IEEEbiography}

\begin{IEEEbiography}[{\includegraphics[width=1in,height=1.25in,clip,keepaspectratio]{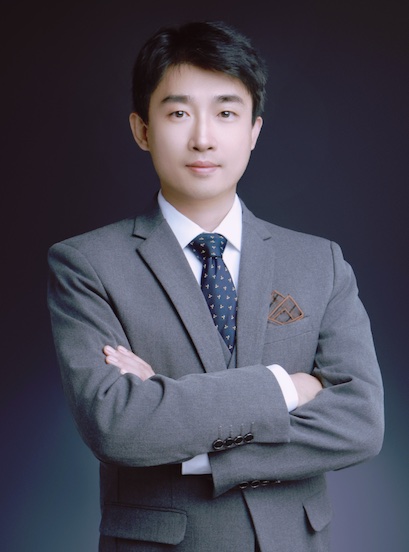}}]{Shuo Wang}  (Senior Member, IEEE) is an Associate Professor at Shanghai Jiao Tong University, specializing in cybersecurity and artificial intelligence. Before joining SJTU, he served as a Senior Research Scientist at the Commonwealth Scientific and Industrial Research Organisation (CSIRO), Australia’s national science agency. He received his Ph.D. in Computer Science from the University of Melbourne in June 2018. Dr. Wang serves as an Associate Editor for IEEE Transactions on Dependable and Secure Computing (TDSC) and IEEE Transactions on Information Forensics and Security (TIFS). His research focuses on Responsible and Trustworthy AI, particularly on enhancing the robustness, interpretability, and reliability of deep learning and graph-based models. He is also dedicated to addressing critical challenges in cybersecurity and privacy across systems, networks, and data-driven infrastructures. Dr. Wang has published more than 60 research papers in prestigious journals and top-tier conferences in information security and artificial intelligence, including IEEE S\&P, NDSS, USENIX Security, CCS, CVPR, ICML, ICLR, IEEE TIFS, IEEE TDSC, IEEE TPDS, IEEE TSC, IEEE TNNLS, WWW, and ESEC/FSE.
\end{IEEEbiography}

\begin{IEEEbiography}[{\includegraphics[width=1in,height=1.25in,clip,keepaspectratio]{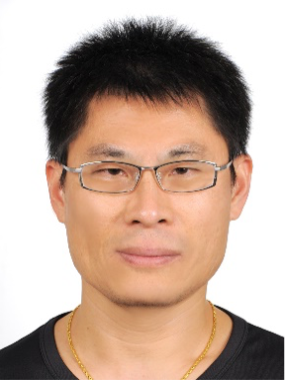}}]{Chun-Shien Lu}
received the Ph.D. degree in Electrical Engineering from National Cheng-Kung University, Tainan, Taiwan in 1998. He is a full research fellow (full professor) in the Institute of Information Science since March 2013 and the executive director in the Taiwan Information Security Center, Research Center for Information Technology Innovation, Academia Sinica, Taipei, Taiwan, since June 2024. His current research interests mainly focus on deep learning, AI security and privacy, and inverse problems. Dr. Lu serves as a Technical Committee member of Communications and Information Systems Security (CIS-TC) and Multimedia Communications Technical Committee (MMTC), IEEE Communications Society, since 2012 and 2017, respectively. Dr. Lu also serves as Area Chairs of ICIP 2019--2026, ICML 2020, ICML 2023--2026, ICLR 2021--2026, NeurIPS 2022--2025, and ACM Multimedia 2022--2026, Senior Area Chair (SAC) of NeurIPS 2026, and Senior program committee of AAAI 2025--2026. Dr. Lu has owned four US patents, five ROC patents, and one Canadian patent in digital watermarking and graphic QR code. Dr. Lu won Ta-You Wu Memorial Award, National Science Council in 2007 and was a co-recipient of a National Invention and Creation Award in 2004. Dr. Lu was an associate editor of IEEE Trans. on Image Processing from 2010/12 to 2014 and from 2018/3 to 2023/6.
\end{IEEEbiography}

\begin{IEEEbiography}[{\includegraphics[width=1in,height=1.25in,clip,keepaspectratio]{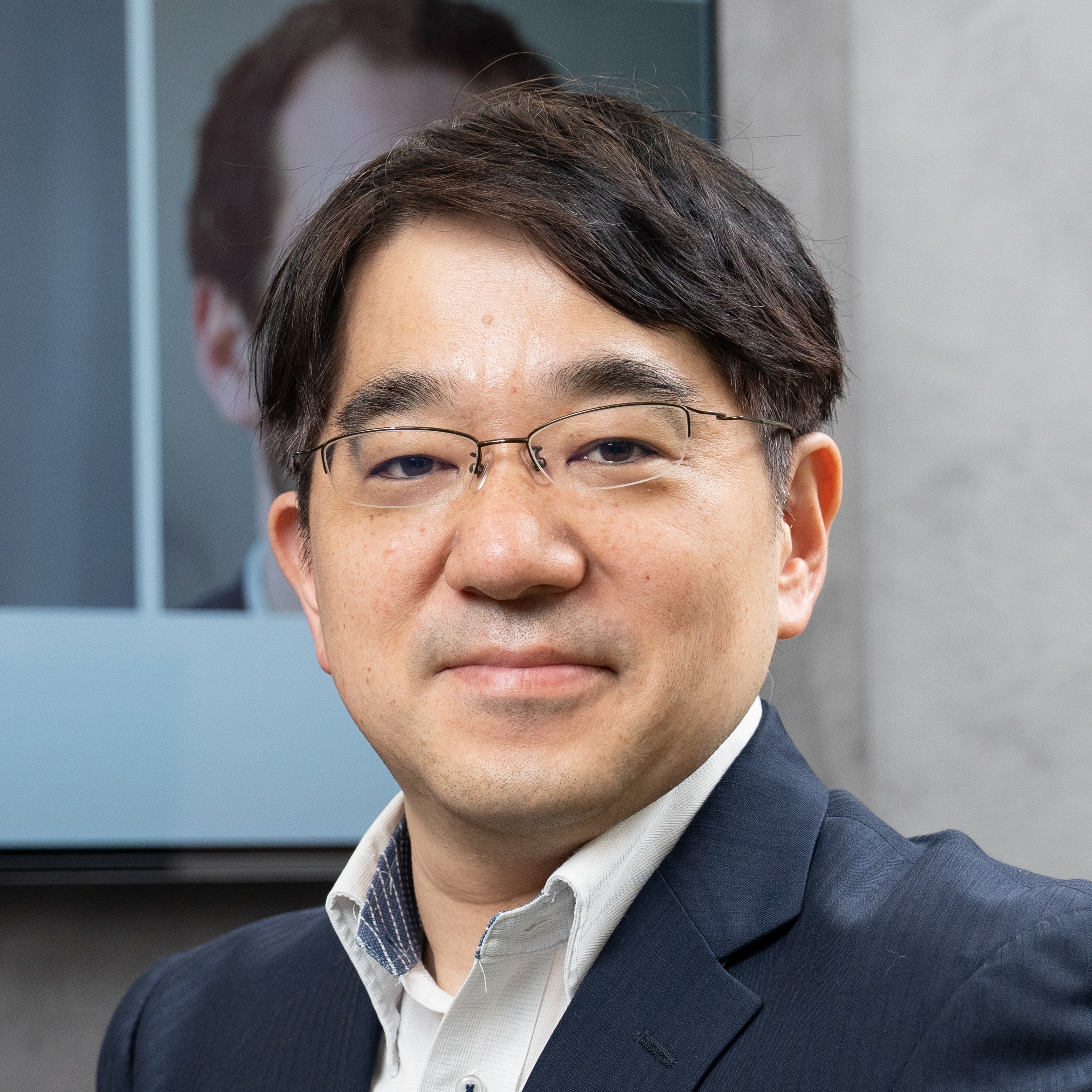}}]{Isao Echizen} (Senior Member, IEEE) received B.S., M.S., and D.E. degrees from the Tokyo Institute of Technology, Japan, in 1995, 1997, and 2003, respectively. He joined Hitachi, Ltd. in 1997 and until 2007 was a research engineer in the company's systems development laboratory. He is currently a director and a professor of the Information and Society Research Division, the National Institute of Informatics (NII), a director of the Global Research Center for Synthetic Media, the NII, a professor in the Department of Information and Communication Engineering, Graduate School of Information Science and Technology, the University of Tokyo, and a professor in the Graduate Institute for Advanced Studies, the Graduate University for Advanced Studies (SOKENDAI), Japan.  He was a visiting professor at the University of Freiburg, Germany, and at the University of Halle-Wittenberg, Germany. He is currently engaged in research on AI security, multimedia security, and multimedia forensics. He is a research director in the CREST FakeMedia project and in the K Program SYNTHETIQ X, Japan Science and Technology Agency (JST). He received the Commendation for Science and Technology by the Minister of Education, Culture, Sports, Science and Technology (Research category) in 2025, the Best Paper Award from the IEICE in 2023, the Best Paper Awards from the IPSJ in 2005 and 2014, the IPSJ Nagao Special Researcher Award in 2011, the DOCOMO Mobile Science Award in 2014, the Information Security Cultural Award in 2016, and the IEEE Workshop on Information Forensics and Security Best Paper Award in 2017. He was a member of the Information Forensics and Security Technical Committee of the IEEE Signal Processing Society. He is the IEICE Fellow, the IPSJ Fellow, the IEEE Senior Member, and the Japanese representative on IFIP and on IFIP TC11 (Security and Privacy Protection in Information Processing Systems), a vice president of APSIPA, and an editorial board member of the IEEE Transactions on Dependable and Secure Computing, the EURASIP Journal on Image and Video Processing, and the Journal of Information Security and Applications, Elsevier.
\end{IEEEbiography}
\clearpage
\appendices

\section{Discussion}
\subsection{Why Inversion is Dangerous and Realistic in Face Recognition?}
Modern face recognition systems, used in authentication, surveillance, and identity verification, increasingly adopt an embedding-based paradigm. Instead of storing raw images, they encode inputs into compact embeddings and compare them via cosine or Euclidean similarity, improving scalability, supporting open-set recognition, and enabling efficient retrieval~\cite{MicrosoftAzureFaceAPI,AmazonRekognition,GoogleCloudVisionAPI,FacePlusPlusAPI}.

\paragraph{Embeddings are assumed to be privacy-preserving}
Because embeddings abstract pixel-level details, they are often viewed as anonymized or non-invertible. Many systems treat them as safe to store, transmit, or expose to clients. However, embeddings retain semantically rich identity features and can be inverted under adversarial conditions.

\paragraph{Model inversion reveals reconstruction feasibility}
Recent studies show that expressive deep models allow embeddings to be used for reconstructing visually realistic, identity-preserving faces~\cite{zhang2024validating}. Such attacks can synthesize high-fidelity headshot-style images from embeddings alone, without identity labels or training data.

\paragraph{Biometric data are non-revocable}
Unlike passwords or cryptographic keys, embeddings encode immutable biometric traits. Once compromised (via client-side memory leaks, inference API exposure, or intermediate model output) they cannot be revoked or re-issued. Even partial leakage enables spoofing, synthetic ID forgery, or impersonation~\cite{boulkenafet2017oulu,tan2021many}.

\paragraph{Embedding leakage is a practical concern}
Many commercial services expose embeddings or similarity scores via SDKs or web APIs. Black-box leakage arises when similarity scores are returned to clients; white-box leakage occurs through insecure deployment, shared inference servers, or side-channel access. Embeddings are often stored in databases for caching, retrieval, or verification, yet these may be compromised via breaches or insider threats (\eg third-party SDKs, optional cloud APIs, misconfigured backups)~\cite{DHSOIG_2020_CBP,Taylor_2019_Guardian_BioStar2_Breach,Burt_2024_BiometricUpdate_Outabox,Thomas_2024_CSOOnline_IndiaLeak,Fortune_2022_Shanghai_Leak,malwarebytes2025logins}. Frequently stored unencrypted and deemed “non-sensitive,” such leaks often go unnoticed, enabling inversion without access to training data or labels.

\textbf{Summary}
Model inversion attacks against embedding-based face recognition are technically feasible and practically relevant. Because embeddings are semantically rich, hard to protect, and irreversible once leaked, they represent a critical security and privacy risk. Mitigating this requires moving beyond the assumption of “embedding-level privacy.”

\subsection{Why Embedding-Based Open-Set Recognition over Closed-Set Classification?}
Modern face recognition systems are fundamentally embedding-based in both research and deployment, driven by advantages over traditional classification-based approaches.

\paragraph{Scalability and generalization to unseen identities}
Embedding-based models enable open-set recognition. Instead of training on a fixed set of classes, models such as FaceNet~\cite{schroff2015facenet} and ArcFace~\cite{deng2019arcface} map facial images into a continuous embedding space, where similarity is computed using metrics like cosine or Euclidean distance. This allows new identities to be enrolled dynamically without retraining.

\paragraph{Industry-wide adoption}
Commercial platforms including Microsoft Azure~\cite{MicrosoftAzureFaceAPI}, Amazon Rekognition~\cite{AmazonRekognition}, and Google Cloud Vision~\cite{GoogleCloudVisionAPI} widely employ embedding-based pipelines. In these systems, embeddings serve as the core representation for matching, verification, and search, often as the final stored artifact tied to user identity.

\textbf{Summary}
Given their central role, embeddings are the primary target in model inversion attacks. Focusing on embeddings aligns with both dominant industry practice and realistic threat models, where raw images are discarded but embeddings are retained for long-term identity recognition.

\subsection{Why Prefer Diffusion Over GAN-based Inversion?}
\paragraph{No truncation bottleneck}
GAN-based inversion, particularly with StyleGAN~\cite{karras2020analyzing}, typically operates in the intermediate latent space $\mathcal{W}$, obtained by mapping standard Gaussian noise from the original latent space $\mathcal{Z}$ through a learned multilayer perceptron (MLP). The disentangled $\mathcal{W}$ space allows more controllable and realistic synthesis. To suppress unnatural outputs, the truncation trick restricts latent codes to a narrow radius around the mean in $\mathcal{W}$. While improving visual quality, truncation reduces expressiveness by excluding low-density regions of the latent space (crucial for reconstructing diverse or atypical identities). In contrast, DDPMs generate images by progressively denoising Gaussian noise without adhering to a fixed latent distribution, enabling unconstrained yet coherent exploration of the generative space and avoiding the realism–diversity trade-off imposed by truncation.

\paragraph{Higher-dimensional, disentangled latent space}
DDPMs operate in a high-dimensional, near-isotropic latent space ($\sim$200K dimensions for $3 \times 256 \times 256$ images), offering far greater flexibility than StyleGAN’s $\sim$512-dimensional $\mathcal{W}$ space. This facilitates fine-grained controls and mitigates the entanglement common in lower-dimensional representations.

\paragraph{Controlled optimization}
Our pipeline integrates on-manifold initialization (robust latent codes), identity-aware warm starts (top $N$ selection), and confidence-aware optimization (ranked adversary with early stopping). These strategies constrain updates to remain within high-probability regions of the diffusion prior, thereby preserving image realism throughout the optimization process.

\textbf{Summary}
A well-controlled DDPM enables precise, artifact-resistant manipulation while maintaining generative realism, surpassing GAN-based methods constrained by truncation and latent entanglement (see evidence in \cref{tab_comparison}).

\subsection{Why Not Naive APGD + DDPM?}

\paragraph{Manifold mismatch}
Diffusion models are trained to map Gaussian noise to the natural image manifold. However, APGD~\cite{croce2020reliable} introduces perturbations that drive latent vectors off this manifold. These deviations are \emph{amplified} during the denoising process, leading to visual artifacts such as halos, patchy textures, and double contours.

\paragraph{Identity-agnostic initialization}
Because the initial noise is sampled independently of the target identity, APGD requires a large perturbation budget to steer the generation toward the desired embedding. This extended traversal increases the likelihood of optimization artifacts.

\paragraph{Lack of transferability}
Naive APGD aggressively overfits to the target model by fully maximizing embedding similarity, even at the cost of introducing more artifacts. As a result, the generated images often fail to generalize to non-target recognition models, reducing cross-model robustness.

\textbf{Summary}
The unconstrained optimization of the naive ``APGD + DDPM'' undermines the diffusion prior that ensures generative realism (see evidence in \cref{tab_more_baseline,fig_compare}).

\subsection{Why We Target Identity Consistency Instead of Pixel-Level Reconstruction?}

A key question in model inversion attacks is whether the objective should be reconstructing the \emph{identical} input image or merely recovering the \emph{identity} of the target. We argue that targeting \textbf{identity-consistent} outputs is both theoretically justified and practically sufficient for evaluating privacy risks.

\paragraph{Privacy leakage centers on identity}
Face recognition systems use embeddings to match identities, not pixel-level details. Thus, reconstructing any image perceived as the same person, even with variations in pose, lighting, or background, constitutes a privacy breach. This reflects realistic threats such as re-identification, impersonation, or spoofing using generated media. The privacy risk arises not from exact replication of an image, but from revealing the individual’s identity.

\paragraph{Theoretical limitations of exact recovery}
Recovering the \emph{exact} input image from its embedding is fundamentally ill-posed. Models such as ArcFace and FaceNet compress high-dimensional facial images into low-dimensional embeddings (e.g., 512D), discarding information irrelevant to identity. This many-to-one mapping means multiple images of the same person collapse to nearly identical embeddings. Exact pixel-wise inversion is therefore unnecessary and, in many cases, \emph{impossible} due to this dimensionality bottleneck.

\paragraph{Practical threat scenario}
This perspective aligns with realistic threat models. In applications such as surveillance, authentication, or forensic recovery, attackers seek to learn or impersonate a person’s identity, not retrieve a specific photo. Privacy-preserving mechanisms should thus be evaluated by their ability to prevent identity inference from embeddings rather than exact image recovery.

\textbf{Summary}
DiffMI adopts identity-preserving reconstruction as its primary objective, consistent with the theory of embedding-based systems and the real-world threats model inversion aims to assess. Our method targets the privacy leakage most relevant to face recognition systems: \emph{who} the person is, not \emph{how} they look in a specific image.

\subsection{Why Compare with These Baselines?}

\paragraph{Two categories of existing methods}
As summarized in \cref{tab_relatedworks}, existing model inversion attacks against face recognition can be grouped into \emph{training-dependent} methods (based on DeconvNet, GANs, or diffusion models) and \emph{training-free} methods (primarily GAN-based, or our proposed diffusion-driven approach).

\paragraph{Primary baseline}
Our main comparison is with MAP$^2$V~\cite{zhang2024validating}, a recent, highly relevant training-free GAN-based attack.

\paragraph{Representative training-dependent methods}
For completeness, we also evaluate representative training-dependent methods based on DeconvNet and GANs~\cite{shahreza2024template,shahreza2024vulnerability}. While not primary baselines due to their substantial training overhead, they serve as benchmarks to highlight the efficiency of training-free approaches.

\paragraph{Naive diffusion-based baseline}
To contextualize our pipeline’s benefits, we evaluate a naive baseline that applies APGD directly to DDPM without our initialization, ranking, or confidence-aware mechanisms. This mimics a straightforward adaptation of GAN-based attacks to the diffusion domain, optimizing latent codes purely via gradient ascent on similarity.

\paragraph{Excluded tasks}
We exclude closed-set classification attacks, which require task-specific training on a fixed set of identities and apply only to seen identities and models, contrary to our focus on unseen cases.

\paragraph{Excluded partial-image attacks}
We also exclude PriDM~\cite{pang2025pridm}, which requires partial target images and thus demands more information than allowed under our threat model.

\textbf{Summary}
Our benchmark selection balances training-free and training-dependent paradigms while adhering strictly to our threat model.

\subsection{Why Our Quantitative Evaluation is Fair?}

\paragraph{Avoiding adversarial overfitting}
A key challenge in evaluating model inversion attacks is avoiding inflated results from overfitting to the target model. Adversarial artifacts may exploit model-specific weaknesses, yielding high similarity scores without preserving meaningful identity features and thus misleading quantitative evaluations~\cite{wang2021similarity,wang2024multi}.

\paragraph{Non-target model evaluation}
To mitigate this, we follow a widely used protocol that measures identity similarity on \emph{non-target} face recognition models not involved in optimization~\cite{zhang2024validating,shahreza2024vulnerability,otroshi2023face,shahreza2024template,dong2023Reconstruct}. This ensures reconstruction success reflects transferable identity features rather than target-specific overfitting. A match across non-target models provides stronger evidence that the underlying identity is faithfully recovered.

\paragraph{Alignment with real-world threats}
This protocol also matches realistic threat scenarios, where attackers aim to reconstruct biometric images capable of impersonating the same identity across different systems. Since biometric traits are immutable, a successful match on non-target models demonstrates a true privacy breach.

\textbf{Summary}
Using non-target models yields a fair and privacy-relevant measure of reconstruction success.

\subsection{Why Model Inversion is Hard to Defend?}

\paragraph{Inference-time leakage is unavoidable}
Model inversion is inherently an inference-time threat. In embedding-based face recognition, attackers need only access to the model’s inputs and outputs (\eg a query image and its embedding) to attempt reconstruction. This reflects common real-world deployments, such as mobile verification or cloud APIs, where embeddings are exposed for identification. As long as a system performs identity matching, an attacker can search for a reconstruction that matches the exposed embedding.

\paragraph{Any high-matching reconstruction is a privacy breach}
For a given embedding, many images can yield high similarity. If reconstruction introduces uncontrolled artifacts, results may be unrecognizable or adversarial; but steering optimization to avoid artifacts often produces an image resembling the true identity. Thus, the ability of face recognition to generalize across pose and illumination inherently enables inversion.

\paragraph{Embedding matching as a proxy for identity}
Embedding similarity correlates strongly with human-perceived identity. Systems are trained to produce similar embeddings for images of the same person despite variation. Therefore, finding an image that closely matches a target embedding recovers key facial characteristics of that person. If a system excels at recognition, it must also be invertible, an inherent tension.

\paragraph{Gradient masking and training-time defenses are insufficient}
Techniques such as gradient hiding or privacy-aware training~\cite{ji2022privacy,mi2023privacy} may reduce overfitting to training data but do not address the core issue: embeddings remain semantically meaningful and exposed at inference. Even without gradient access, query-based optimization can recover facial structure.

\paragraph{Face recognition accuracy implies invertibility}
Paradoxically, the better a system is at identity recognition, the more vulnerable it is to inversion. If a model can reliably distinguish individuals, inverting its output can produce an image that must resemble the underlying person. This stems from the design of embedding spaces to preserve identity semantics.

\textbf{Summary}
Model inversion is difficult to defend because it exploits core design principles of face recognition. As long as embeddings are exposed for matching and preserve identity in feature space, attackers can reconstruct the underlying identity. Mitigation requires rethinking how embeddings are generated and shared, rather than assuming safety from the absence of raw images.

\begin{table*}[!t]
    \footnotesize
    \centering
    \begin{threeparttable}
        \caption{Key notations and their corresponding definitions.}
        \label{tab_notation}
        \setlength{\tabcolsep}{4.9mm}{\begin{tabular}{cllc}
            \toprule
            Notation&Definition&Remark&Reference\\
            \midrule
            $F(\cdot)$&Embedding function (Face recognition)&White- or black-box knowledge&\multirow{17}{*}{\cref{problem_loss}}\\
            $x$&Facial image&&\\
            $z$&Feature Embedding&&\\
            $S(\cdot,\cdot)$&Similarity function (cosine)&&\\
            $\tau_F$&Similarity decision threshold&&\\
            $x^\mathrm{tgt}$&Target facial image&Unknown to Attackers&\\
            $z^\mathrm{tgt}$&Target embedding, transformed from the target face&Known to Attackers&\\
            $\hat{x}$&Reconstructed image&Attack Output&\\
            $\hat{z}$&Feature embedding of the reconstructed image&&\\
            $G(\cdot)$&Generative function (DDPM)&Pretrained, unconditional&\\
            $x_G$&Latent code, drawn from a random Gaussian distribution&Attack Input\\
            $x'_G$&Manipulated latent code&&\\
            $\delta$&Adversarial perturbations on the latent code&&\\
            $\|\cdot\|_p$&$L_p$-norm&&\\
            $\epsilon$&Perturbation magnitude&Attack Setting&\\
            $\mathcal{L}$&Objective function&&\\
            $\tau_C$&Confidence threshold (sufficient similarity)&Attack Setting&\\
            \midrule
            $K(\cdot)$&Gaussian normality test function ($K^2$ test)&&\multirow{3}{*}{\cref{ktest}}\\
            $p_K$&Gaussian normality (the probability of following a normal distribution)&&\\
            $\tau_K$&Threshold of Gaussian normality&Attack Setting&\\
            \midrule
            $D(\cdot)$&Face detection function (MTCNN)&&\multirow{3}{*}{\cref{face_detection}}\\
            $p_D$&Face detection confidence score&&\\
            $\tau_D$&Threshold of detection confidence&Attack Setting&\\
            \midrule
            $V$&Volume of reliable latent codes (Step (a))&Attack Setting&\multirow{3}{*}{\cref{TopN}}\\
            $N$&Top $N$ selection (Step (b))&Attack Setting&\\
            $Q$&Query Efficiency&&\\
            \midrule
            $t_{max}$&Maximum iterations per adversarial attack&Attack Setting&\multirow{3}{*}{\cref{manipulaiton}}\\
            $Q_{max}$&Maximum number of queries (only for black-box attacks)&Attack Setting&\\
            $\lfloor \cdot \rfloor$&Floor function&&\\
            \midrule
            $\mathbb{I}(\cdot)$&Indicator function&&\multirow{4}{*}{\cref{settings}}\\
            $I$&Total number of attack samples&&\\
            $x^{j \neq \mathrm{tgt}}$&Facial images distinct from the target, associated with the same identity&&\\
            $J$&Total number of $x^{j \neq \mathrm{tgt}}$ for each identity&&\\
            \bottomrule
        \end{tabular}}
    \end{threeparttable}
\end{table*}

\section{Notations}
The key notations and their corresponding definitions are summarized in \cref{tab_notation}. 

\section{Model Architectures and Datasets}
We identify that training-dependent model inversion attacks incur high computational cost and limited generalizability due to their reliance on target-specific generator training. To address these limitations, we propose DiffMI, which leverages a fixed, unconditional diffusion model (DDPM) to attack unseen target identities and face recognition models without retraining.

For a comprehensive evaluation, we consider diverse model architectures and datasets, as summarized in \cref{tab_diversity}. The DDPM generator and the four recognition models differ substantially in both architecture and training data. Our primary evaluation uses LFW, which is excluded from training for both the generator and recognition models, ensuring all cross-model evaluations involve architectural and distributional mismatches.

We further investigate whether alignment between the generator’s training data and the attack domain improves reconstruction. This simulates a practical case where the attacker’s data distribution resembles the target’s (\eg ID photos often share similar format and content). To this end, we select 1,000 target identities from CelebA-HQ (the dataset used to train DDPM) and compare results with LFW targets. As shown in \cref{tab_crossmodel}, prior alignment yields only marginal improvements or even slight degradation in some cases (\eg +2.82\% accuracy on LFW \vs CelebA-HQ for DCTDP). These results suggest that, in training-free open-set inversion, alignment with the generator’s training distribution offers limited practical benefit.

Lastly, we examine whether our evaluation dataset size (1,000 images per dataset) is sufficient for a training-free method. Results in \cref{tab_datasetsize} show that performance stabilizes once the dataset size reaches 500 samples, as the change from 500 to 1,000 is only marginal. Therefore, as DiffMI is training-free, evaluation variance is largely independent of dataset size.

\begin{table}[!t]
    \footnotesize
    \centering
    \begin{threeparttable}
        \caption{Impact of dataset size on evaluation variance.}
        \label{tab_datasetsize}
        \setlength{\tabcolsep}{4.3mm}{\begin{tabular}{ccccc}
            \toprule
            Size&Type I (\%)&Gap&Type II (\%)&Gap\\
            \midrule
            \ \ \ 100&98.00&\multirow{2}{*}{1.00}&91.59&\multirow{2}{*}{2.77}\\
            \ \ \ 200&99.00&\multirow{2}{*}{0.45}&94.36&\multirow{2}{*}{0.22}\\
            \ \ \ 500&98.55&\multirow{2}{*}{0.00}&94.58&\multirow{2}{*}{0.14}\\
            1,000&98.55&&94.72&\\
            \bottomrule
        \end{tabular}}
        \begin{tablenotes}
            \item Target model: ArcFace~\cite{deng2019arcface}; evaluation dataset: LFW~\cite{hua2008labeled}.
        \end{tablenotes}
    \end{threeparttable}
\end{table}



\end{document}